\documentclass{aa}  

\usepackage{graphicx}
\usepackage{txfonts}
\usepackage{comment}
\begin{document}

   \title{Mitigating stellar activity jitter with different line lists for least-squares deconvolution: analysis of a parametric and a randomised line selection}
   \titlerunning{Mitigating stellar activity with different line selections}

   \author{S. Bellotti \inst{1}
          \and
          P. Petit \inst{1}
          \and
          J. Morin \inst{2}
          \and
          G. A. J. Hussain \inst{3}
          \and
          C. P. Folsom \inst{4,5}
          \and
          A. Carmona \inst{6}
          \and
          X. Delfosse \inst{6}
          \and
          C. Moutou \inst{1}
          }
   \authorrunning{Bellotti et al.}
    
   \institute{
            Institut de Recherche en Astrophysique et Plan\'etologie,
            Universit\'e de Toulouse, CNRS, IRAP/UMR 5277,
            14 avenue Edouard Belin, F-31400, Toulouse, France\\
            \email{stefano.bellotti@irap.omp.eu}
        \and
             Laboratoire Univers et Particules de Montpellier,
             Universit\'e de Montpellier, CNRS,
             F-34095, Montpellier, France
        \and
             Science Division, Directorate of Science, 
             European Space Research and Technology Centre (ESA/ESTEC),
             Keplerlaan 1, 2201 AZ, Noordwijk, The Netherlands
        \and
            Department of Physics \& Space Science,
            Royal Military College of Canada,
            PO Box 17000 Station Forces, Kingston, ON, Canada K7K 0C6
        \and 
            Tartu Observatory, 
            University of Tartu, 
            Observatooriumi 1, Tõravere, 61602 Tartumaa, Estonia
        \and
            Universit\'e Grenoble Alpes, CNRS, IPAG, 38000 Grenoble, France
             }
   \date{Received ; accepted }

 
  \abstract
   {Stellar activity limits the radial velocity (RV) search and characterisation of exoplanets, as it introduces spurious noise (called ``jitter'') in the data sets and prevents the correct retrieval of a planetary signal. This is key for M dwarfs, considering that they manifest high activity levels and are primary targets for present and future searches of habitable Earth-like planets. To perform precise RV measurements, multi-line numerical techniques like cross-correlation and least-squares deconvolution (LSD) are typically employed.}
   {Effective filtering of activity is crucial to achieve the sensitivity required for small planet detections. Here, we analyse the impact of selecting different line lists for LSD on the dispersion in our RV data sets, to identify the line list that most effectively reduces the jitter.}
   {We employ optical spectropolarimetric observations of the active M dwarf EV Lac collected with ESPaDOnS and NARVAL, and study two line downselection approaches: a parametric one based on line properties (depth, wavelength, magnetic sensitivity) and a randomised algorithm that samples the line combination space. We test the latter further to find the line list that singles out the activity signal from other sources of noise, and on AD Leo and DS Leo to examine its consistency at mitigating jitter for different activity levels. The analysis is complemented with planetary injection tests.}
   {The parametric selection yields a RV RMS reduction of less than 10\%, while the randomised selection a systematic $>$50\% improvement, regardless of the activity level of the star examined. Furthermore, if activity is the dominant source of noise, this approach allows the construction of lists containing mainly activity-sensitive lines, which could be used to enhance the rotational modulation of the resulting data sets and determine the stellar rotation period more robustly. Finally, the output line lists allow the recovery of a synthetic planet (0.3-0.6 M$_\mathrm{Jup}$ on a 10 d orbit) in the presence of both moderate (20 m s$^{-1}$ semi-amplitude) and high (200 m s$^{-1}$) activity levels, without affecting the planet signal (between 60 and 120 m s$^{-1}$) substantially.}
   {} 


   \maketitle

%

\section{Introduction}\label{sec:intro}

A key scientific objective of exoplanetary sciences is to find and characterise habitable Earth-like planets, scrutinise their atmosphere and search for potential biomarkers; it is the main driver for current space-based missions such as TESS and CHEOPS \citep{Ricker2015,Benz2020}, as well as for future ones like JWST, PLATO, and ARIEL \citep{Gardner2006,Rauer2014,Tinetti2018} and ground-based facilities like ESO ELT \citep{Marconi2021}. RV measurements are fundamental to infer physical properties such as mass and density of the planets detected by the transit method, therefore providing a deeper characterisation.

The success of velocimetric surveys relies both on the instrumental precision and the ability of numerical techniques to remove the effects of stellar activity from the data sets. Considering that modern instruments such as ESPRESSO \citep{Pepe2013} and EXPRES \citep{Jorgenson2016} reach a precision of a few dozens of cm s$^{-1}$, i.e. just above the one required by a habitable Earth-like planet around a Sun-like star, the remaining fundamental limitation is the activity-induced variability \citep{Meunier2021}. Stellar activity produces surface inhomogeneities that distort the spectral line profiles and result in spurious Doppler shifts, reaching a few m s$^{-1}$ in solar-like stars \citep{Dumusque2021,CollierCameron2021} and over 1 km s$^{-1}$ for, e.g., young (Myr-old) T Tauri stars \citep{Donati2016}. The RV signal of a habitable Earth-like planet is consequently drowned and its detection severely hampered.

M dwarfs are attractive targets for low-mass planet detections: they outnumber the other stars in the solar neighbourhood, have high Earth-like planet occurrence rates \citep{Kopparapu2013, Gaidos2016}, and have closer habitable zones (implying magnified RV reflex motion and higher transit probability). Furthermore, the habitable zone of M dwarfs being more compact, the orbital period of a habitable planet cannot be mistaken with long-term evolution of the stellar magnetic cycle, contrarily to active Sun-like stars \citep{Meunier2010}. However, M dwarfs exhibit high levels of activity and that persist for Gyr timescales \citep{West2008}, hence robust activity-filtering techniques are essential to ensure precise RV measurements and reliable planet detections. 

RV measurements are commonly performed with numerical techniques such as cross correlation \citep{Baranne1996,Pepe2002}, least-squares deconvolution (LSD, \citealt{Donati1997,Kochukhov2010}) and least-squares template matching \citep[e.g.,][]{Anglada2012,Astudillo2015,Zechmeister2018}. The latter technique gives precise results for M dwarfs, because it makes use of a master spectrum template with high signal-to-noise ratio (SNR) and with a richer RV content than a synthetic spectrum with a lower number of absorption lines. However, both cross correlation and LSD produce a high-SNR line profile from which we can compute the FWHM and bisector. These are known to correlate with stellar activity \citep{Queloz2001,Boisse2009,Queloz2009} and can therefore be used to disentangle activity signatures from genuine planetary signals. A comparison of the performance between cross-correlation and LSD will be investigated in a forthcoming paper (Bellotti et al., in prep.).

\citet{Dumusque2018} and \citet{Cretignier2020} employed the fact that spectral lines have different sensitivity to activity \citep[e.g.,][]{Davis2017,Lisogorskyi2019} and designed a line-by-line RV extraction method for $\alpha$ Cen B with which the activity signal is either enhanced or mitigated by a factor of two. The precision with which to robustly identify individual lines is however limited by blends and photon noise, an aspect that becomes more relevant when considering cooler stars, since they are fainter and feature dense spectra. So far, no line-by-line method has been applied to M dwarfs in the visible domain, as its extension is not straightforward. 

In this paper, we analyse different line selection approaches to build synthetic line lists (called ``masks'' throughout) for least-squares deconvolution, with the intention to develop a new activity-mitigating procedure tailored for active M-dwarfs observed in the optical. Its extension to the near-infrared regime is also planned (Bellotti et al., in prep.). We test a parametric selection based on line properties (depth, wavelength and magnetic sensitivity) and a randomised selection that samples the line combination space. We assess the extracted subsets of lines by looking at the resulting dispersion of the RV data sets, in order to find a mask that reduces the activity jitter. Overall, our approach benefits from the multi-line nature of LSD, since it produces a high-SNR profile cleared from line blends. 

The paper is structured as follows. In Sec.~\ref{sec:data}, we describe the spectropolarimetric data sets that have been used in this study. Sec.~\ref{sec:parametric} and Sec.~\ref{sec:randomized} are dedicated to the description of the parametric and randomised selection used to create the line subsets, respectively, and in Sec~\ref{sec:conclusions} we present our conclusions.
\section{Data sets}\label{sec:data}

The analysis presented here focuses mainly on EV Lac (GJ 873), an active M4.0 star whose magnetic field properties have been extensively studied \citep[e.g.,][]{Morin2008,Shulyak2019}. This star has a rotation period of 4.38 days and is notoriously active, with an X-ray-to-bolometric luminosity ratio ($\log(\mathrm{L}_\mathrm{X}/\mathrm{L}_\mathrm{bol})$) of -3.10 \citep{Wright2011} and a CaII H\&K chromospheric activity index ($\mathrm{logR'}_\mathrm{HK}$) of -3.75 \citep{BoroSaikia2018,Noyes1984}. The high activity level of EV Lac makes it an optimal target for our study, because it is the dominant source of noise in the RV data sets and allows us to compare the performance of different line selection approaches. 

\begin{figure}[t]
    \centering
    \includegraphics[width=\columnwidth]{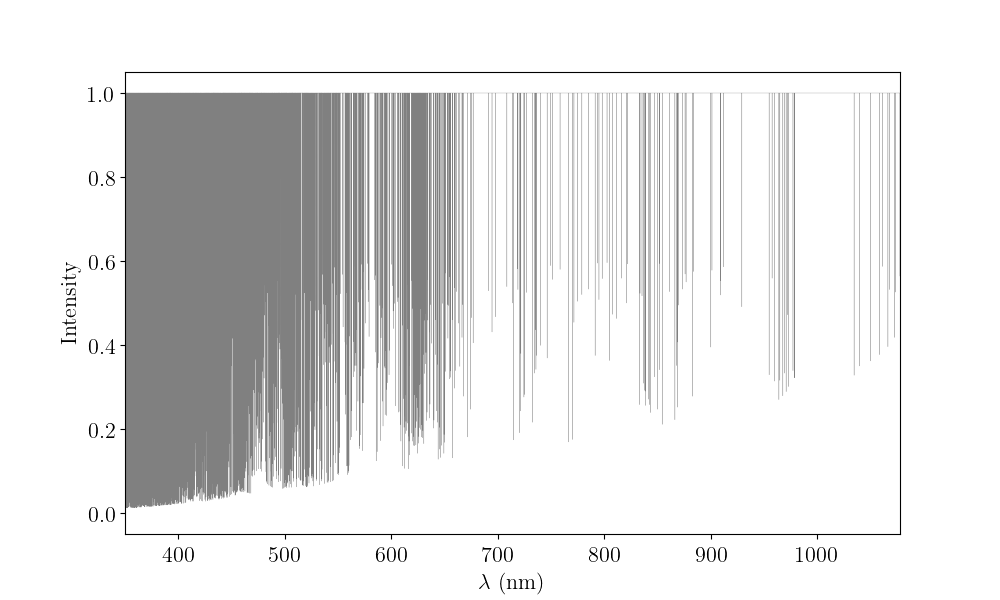}
    \caption{Numerical mask generated with a LTE model atmosphere from VALD showing atomic lines for M dwarfs like EV Lac.}
    \label{fig:mask}%
\end{figure}

We use 57 spectropolarimetric observations in the visible collected between 2005 and 2016 with the twin spectropolarimeters ESPaDOnS on the 3.6-m Canada–France–Hawaii Telescope (CFHT) located atop Mauna Kea in Hawaii, and NARVAL on the 2-m Télescope Bernard Lyot (TBL) at the Pic du Midi Observatory in France \citep{donati2003}. The continuum normalised spectra were retrieved from PolarBase \citep{Petit2014}.

We derive Stokes $I$ (unpolarised) and $V$ (circularly polarised) profiles via LSD. This numerical technique combines the information from thousands of spectral lines to obtain a mean high signal-to-noise ratio (SNR) line profile \citep{Donati1997,Kochukhov2010}. With LSD, the spectrum is regarded as the convolution between a mean profile and a line mask; the latter is basically a weighted Dirac comb reproducing the arrangement of lines in the stellar spectrum, with the associated wavelengths, depths and Land\'e factors (i.e. sensitivities to the Zeeman effect at a given wavelength). Compared to a cross-correlation, LSD removes the auto-correlation function of the mask to provide a cleaner mean profile.

For stars like EV Lac, we employ the same line mask used in \citet{Morin2008}. It is generated using the Vienna Atomic Line Database (VALD) with a local thermodynamic equilibrium model \citep{Gustafsson2008} and corresponds to $T_{\mathrm{eff}}=3500$ K, $\log g=$ 5.0 [cm s$^{-2}$] and $v_{\mathrm{micro}}=$ 2 km s$^{-1}$, and contains 4216 atomic lines between 350--1080 nm (Fig.~\ref{fig:mask}) with depths larger than 40\% the continuum level. The number of lines in range for our observations is 3300. This will be referred to as ``full mask'', while the extracted subsets as ``sub-masks''. Note that to extract precise radial velocities, an empirical mask with a rich content of spectral lines is generally used, rather then a synthetic one.

To compute the RV values, we simultaneously fit a Voigt function and a linear continuum to each Stokes $I$ profile, within a $\pm$ 20 km s$^{-1}$ velocity interval from the profile centre. The additional linear continuum fit is a precaution to prevent biases of the RV measurement due to an imperfect spectrum normalisation. In addition, we compute the longitudinal magnetic field
\begin{equation}
\mathrm{B}_l\;[G] = \frac{-2.14\cdot10^{11}}{\lambda_0 \mathrm{g}_{\mathrm{eff}}c}\frac{\int vV(v)dv}{\int(I_c-I)dv} \,,
\label{eq:Bl}
\end{equation}
where $\lambda_0$ and $\mathrm{g}_\mathrm{eff}$ are the normalisation wavelength and Land\'e factor of the LSD profiles, $I_c$ is the continuum level, $v$ is the radial velocity associated to a point in the spectral line profile in the star's rest frame and $c$ the speed of light in vacuum \citep{Donati1997}. B$_l$ measurements are performed within a velocity interval of $\pm$ 30 km s$^{-1}$, to include the absorption ranges of both Stokes $I$ and $V$ profiles.

\begin{figure}[t]
    \centering
    \includegraphics[width=\columnwidth]{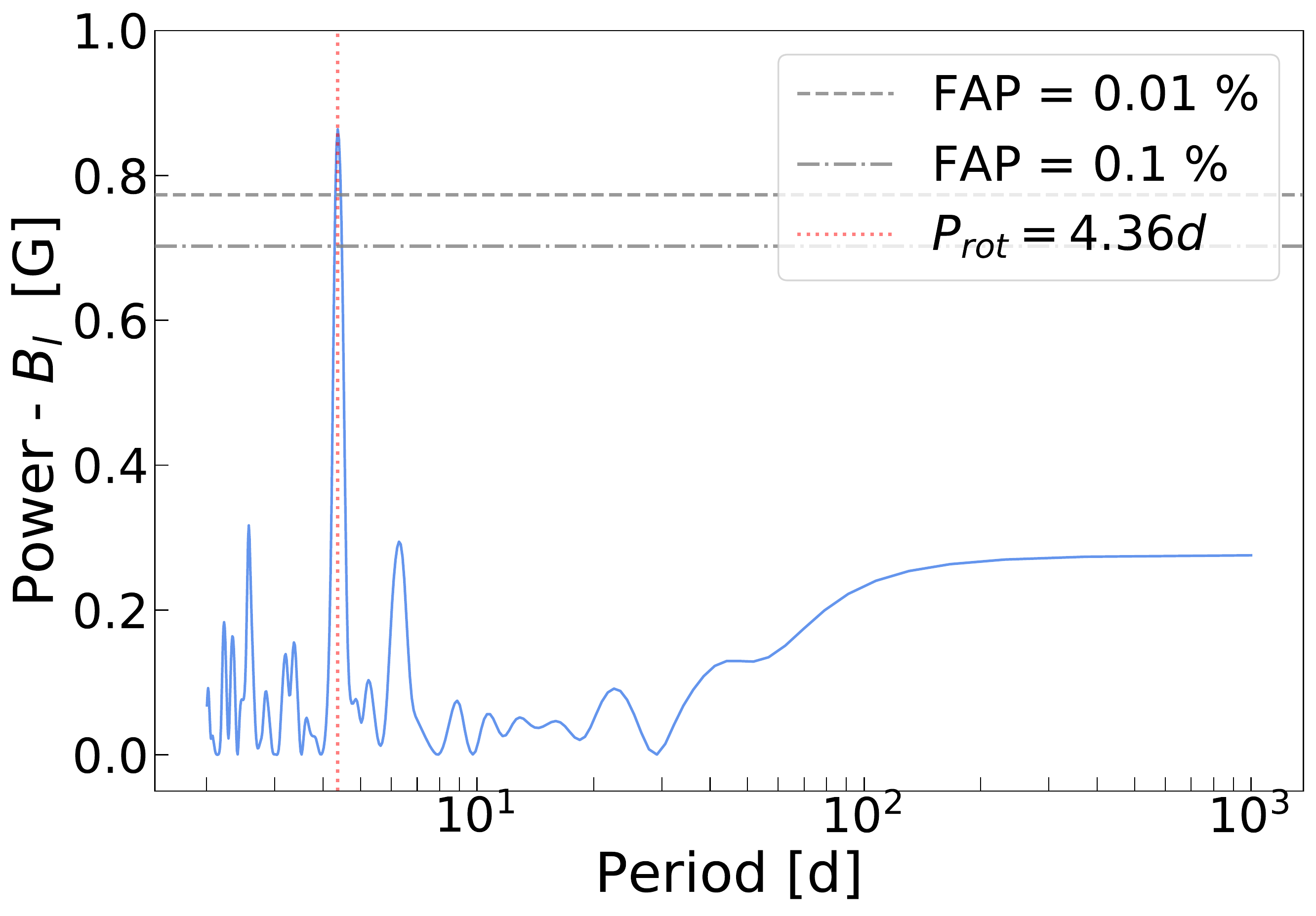}
    \caption{Generalised Lomb-Scargle periodogram of the longitudinal magnetic field of EV Lac, for the 2010 data set (20 observations). The data set is preliminarily detrended with a quadratic fit in order to remove long-term variations. The normalisation of the periodogram power and the false alarm probability (FAP) levels are computed with the prescription in \citep{Zechmeister2009}. We find a significant (FAP$<$ 0.011\%) peak at a rotation period of 4.36 $\pm$ 0.05 d, which is consistent with previous studies.}
    \label{fig:LSPdefault}%
\end{figure}

B$_l$ is a known magnetic activity tracer, as its temporal variation is modulated at the stellar rotation period. Both \citet{Folsom2016} and \citet{Hebrard2016} showed that it performs better than, e.g., RVs at retrieving the stellar rotation period, especially because it is not expected to be influenced by the presence of a planet. We therefore carry out a preliminary period analysis on the B$_l$ 2010 data set using the generalised Lomb-Scargle (GLS) periodogram \citep{Zechmeister2009,Czesla2019}, and find a significant peak at a rotation period of 4.36 $\pm$ 0.05 d (Fig.~\ref{fig:LSPdefault}), consistent with previous studies \citep[e.g.,][]{Wright2011,Morin2008}. This value will be used in the following analysis.

In the following sections, the performance of the sub-masks will be quantified primarily by the dispersion of the associated data sets (both for RV and B$_l$). We will indeed use ``stable'' or ``unstable'' to identify sub-masks leading to an improvement or a degradation in RMS relative to the full mask case, respectively. 
We will also perform contemporaneous Monte Carlo simulations to determine the RV precision associated to each sub-mask and use it as a metric to discern whether the line selection is photon noise limited. More precisely, for each time series examined, we (1) select the spectrum corresponding to the Stokes $I$ profile with SNR closest to the SNR of the median Stokes $I$ profile, (2) inject white noise into the spectrum in the form $\mathcal{N}(0,\sigma_\mathrm{i})$(with $\sigma_\mathrm{i}$ the error bar for each data point of the chosen spectrum), (3) apply LSD with the examined sub-mask and (4) measure RV. The RV RMS of 100 iterations will be representative of the photon noise level associated to the sub-mask. Given that spectra are collected individually, it will not be surprising if some estimates of the photon noise are lower than the stability of the instrument (30 m s$^{-1}$, \citealt{Moutou2007}).
\section{Parametric selection}\label{sec:parametric}

For this first approach, the full mask is cleaned of lines close to H$\alpha$ or belonging to telluric windows, i.e. in the ranges [627,632], [655.5,657], [686,697], [716,734], [759,770], [813,835], and [895,986] nm; hence we conservatively use 3240 lines in total. The RV time series obtained using the full mask and all the observations has a RMS dispersion of 165 m s$^{-1}$ and a RV precision of 8 m s$^{-1}$. 

\begin{figure*}[t]
    \centering
    \includegraphics[width=\linewidth]{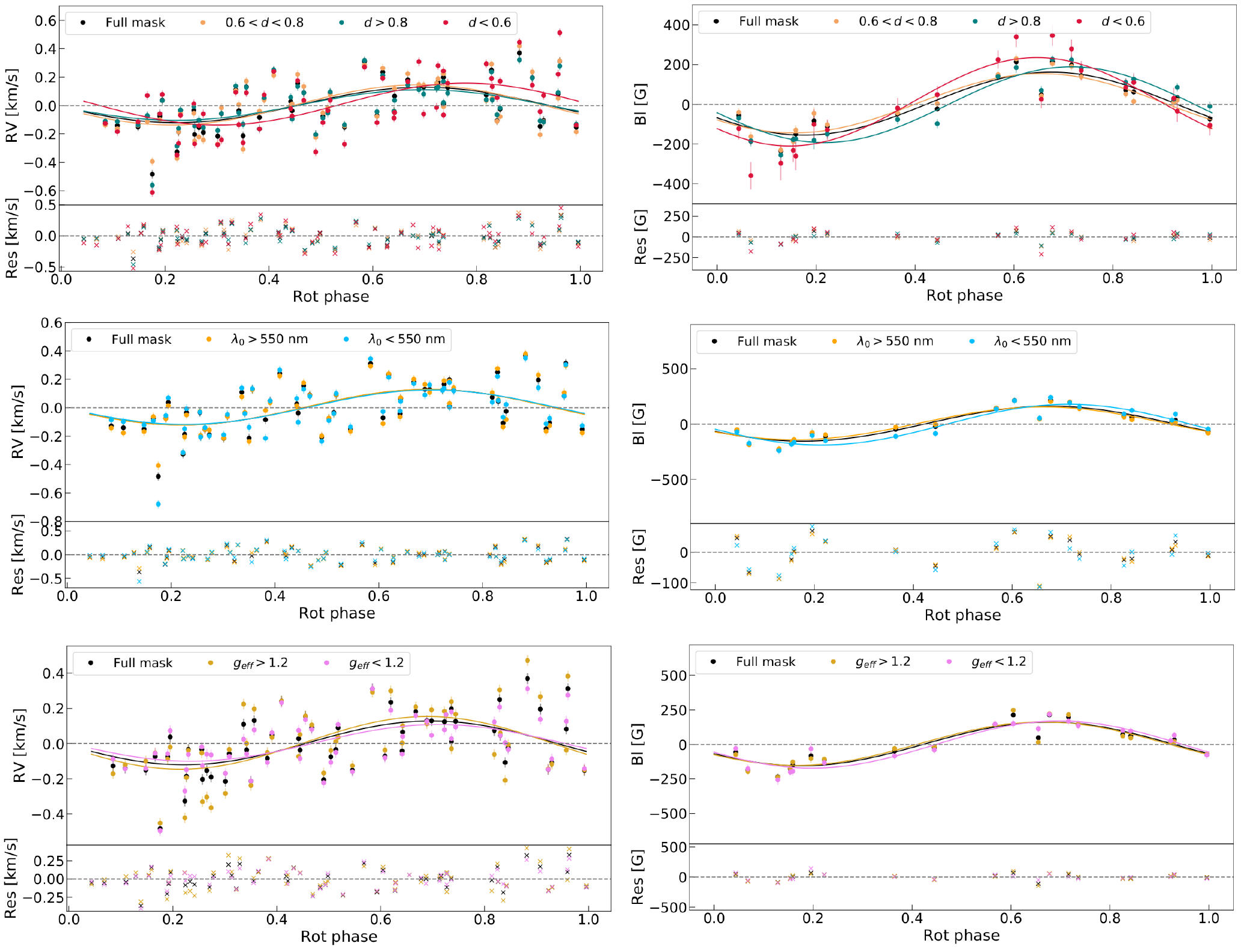}
    \caption{Comparison of RV and B$_l$ data sets obtained with a parametric selection of sub-masks. Top: RV (left) and B$_l$ (right) data sets obtained with a depth selection. Middle: same with a wavelength selection. Bottom: same with a Land\'e factor selection. In all panels, black data points correspond to either the RV or the B$_l$ values obtained with the full mask, whereas the continuous lines represent the sinusoidal fit of the associated data set. The mean of the data set is subtracted to allow a simpler comparison. For RV estimates, we set the error bars to 30 m s$^{-1}$ as derived from the telluric lines-based wavelength calibration of the spectra \citep{Moutou2007}, which is larger than the photon noise for each of these data sets.
    For B$_l$ we use the formal uncertainties. The displayed data sets are 3-$\sigma$ clipped to prevent outliers from affecting the results. For B$_l$ we display the densest epoch (i.e. 2010).
    \label{fig:parametric}}%
\end{figure*}

The line selection consists in splitting the full mask based on the following line parameters: depth ($d$), wavelength ($\lambda$) and Land\'e factor (g$_{\mathrm{eff}}$). The parametric threshold is set so that the SNR distribution of the observed Stokes $V$ profiles for the two sub-masks is compatible; this way we ensure the same SNR and a consistent comparison of the effects of different sub-masks. For each sub-mask, we perform a three-parameters (semi-amplitude $K$, phase correction and offset) Levenberg-Marquardt least squares sinusoidal fit to both the RV and B$_l$ data sets, assuming the stellar rotation period obtained in Sec.~\ref{sec:data}. The inclusion of higher order harmonics of the stellar rotation period \citep{Boisse2011} does not lead to a substantial improvement in the results throughout. The results are illustrated in Fig.~\ref{fig:parametric} and summarised in Table~\ref{tab:parametric} for different parametric selections.

\begin{table*}
\caption{Summary of the RV and B$_l$ data sets characteristics. The columns are: (1) the parametric criteria used, (2) the number of lines in the sub-mask, (3) the RMS of the data set, (4) the RV precision from the MC simulations, (5) the semi-amplitude of the sinusoidal fit to the data set, (6) the reduced $\chi^2$ (three fit parameters and 57 observations for RV and 20 observations for B$_l$), and (7) the RMS of the fit residuals. The reported values for B$_l$ refer to the epoch with most observations (2010), which was used in Sec.~\ref{sec:data}, but the results are analogous also for 2007 (the second most dense data set).}          
\label{tab:parametric}     
\centering                       
\begin{tabular}{c c c c c c c}      
\hline\hline                 
Selection & n$_\mathrm{lines}$ & RMS & Precision & $K$ & $\chi^2_\nu$ & RMS$_\mathrm{res}$\\   
\hline
\multicolumn{7}{c}{Radial Velocity}\\
 & & [m s$^{-1}$] & [m s$^{-1}$] & [m s$^{-1}$] & & [m s$^{-1}$]\\
\hline                    
   Full                 & 3240 & 167 & 5 & 124$\pm$26 & 23.2 & 140  \\
   $d<$0.6              & 1058 & 208 & 22 & 147$\pm$32 & 36.6 & 176  \\    
   0.6$<d<$0.8          & 926  & 183 & 10 & 141$\pm$28 & 27.1 & 152  \\
   $d>$0.8              & 1231 & 157 & 8 & 107$\pm$25 & 22.0 & 136  \\
   $\lambda>$550 nm     & 314  & 168 & 8 & 125$\pm$26 & 23.6 & 141  \\
   $\lambda<$550 nm     & 2872 & 173 & 7 & 121$\pm$27 & 26.2 & 149  \\ 
   g$_\mathrm{eff}>$1.2 & 1649 & 199 & 8 & 149$\pm$31 & 32.9 & 167  \\
   g$_\mathrm{eff}<$1.2 & 1495 & 151 & 7 & 104$\pm$24 & 20.0 & 130  \\ 
\hline
\multicolumn{7}{c}{Longitudinal Magnetic Field}\\
 & & [G] & [G] & [G] & & [G]\\
\hline                    
   Full                 & 3240 & 133 & $\dots$ & 158$\pm$15 & 10.6 & 48 \\
   $d<$0.6              & 1058 & 200 & $\dots$ & 222$\pm$26 & 30.9 & 82 \\    
   0.6$<d<$0.8          & 926  & 129 & $\dots$ & 149$\pm$16 & 12.7 & 53 \\
   $d>$0.8              & 1231 & 152 & $\dots$ & 190$\pm$15 & 10.1 & 46 \\
   $\lambda>$550 nm     & 314  & 128 & $\dots$ & 148$\pm$15 & 10.4 & 48 \\
   $\lambda<$550 nm     & 2872 & 146 & $\dots$ & 184$\pm$15 & 11.6 & 50 \\ 
   g$_\mathrm{eff}>$1.2 & 1649 & 137 & $\dots$ & 157$\pm$17 & 14.8 & 55 \\
   g$_\mathrm{eff}<$1.2 & 1495 & 138 & $\dots$ & 170$\pm$14 & 8.9  & 51 \\ 
\hline                                 
\end{tabular}
\end{table*}

\subsection{Depth case}\label{sec:param_depth}

We select lines with depth lower than 0.6, between 0.6 and 0.8, and larger than 0.8, resulting in three sub-masks of 926, 1231, and 1058 lines, respectively. For RV, the RMS dispersion is decreased only in the $d>$0.8 case by 6\%, while it is deteriorated up to 25\% in the other cases. The semi-amplitude follows a similar trend, but with a more moderate deterioration in the $d<$0.6 case. Depending on the sub-mask considered, the RV data sets show systematic shifts up to 100 m s $^{-1}$ relative to the full-mask data set, likely due to a different line sensitivity to convective motions and corresponding velocity fields \citep{Gray2005,Morgenthaler2012,Beeck2013,Meunier2017}. Another reason could be instrumental systematics affecting specific spectral lines \citep{Cretignier2021}.


For B$_l$, no significant improvement in either RMS or semi-amplitude is observed. The $d<$0.6 case yields the least precise estimates, resulting in deviant magnetic field measurements (with respect to the full mask) and a poorer sinusoidal fit (increased $\chi^2_\nu$). This suggests that deeper lines have a stabilising effect on the time series (also given their higher effective SNR) in agreement with \citet{Reiners2016} and \citet{Cretignier2020}. 


\subsection{Wavelength case}

Activity signatures induced by spots are strongly chromatic \citep{Reiners2010}, hence a wavelength selection may help locate spectral regions with reduced RV dispersion. We select lines with wavelength above and below 550 nm, resulting in two sub-masks of 314 and 2872 lines, respectively. The RV values exhibit a loss of precision up to 3.6\% for both sub-masks, and the B$_l$ values feature a negligible $<$ 4\% improvement in RMS only when using red ($>$ 550 nm) lines. Likewise, the semi-amplitude of the sinusoidal fit for both quantities shows either a degradation or marginal improvement. Note that despite the different number of lines in the sub-masks (due to the larger SNR in the red part of the spectrum), the RV and B$_l$ estimates are highly compatible with the full mask values.

\subsection{Land\'e factor case}
An additional line parameter to examine is the Land\'e factor, considering that \citet{Johns-Krull1996} reported measurable Zeeman broadening in magnetic-sensitive optical lines for EV Lac. We select lines with g$_{\mathrm{eff}}$ above and below 1.2, resulting in two sub-masks of 1649 and 1495 lines, respectively. The RV dispersion (semi-amplitude) shows a $<$ 10\% (16\%) improvement when using magnetically insensitive lines and a 20\% (21\%) degradation with the opposite sub-mask; the results for B$_l$ are similar to the full mask with a slight deterioration with both sub-masks, and the values are reasonably consistent within error bars. We also tried to use the product $\lambda_0\cdot g_\mathrm{eff}$ as selection criterion, to have a more accurate measure of the magnetic susceptibility, but the result did not change.

\subsection{Summary of parametric selection results}
We also extended the initial mask to contain lines as shallow as 10\%. With 9470 lines in total, we applied similar parametric thresholds (the depth case was modified to ensure same SNR) to investigate any benefit of the substantial increase in the number of lines. We found similar trends to the previous analysis, with a general degradation of the precision and of B$_l$ reliability, confirming that very shallow lines should be rejected when defining line masks.

From the results of this first approach, we conclude that building a sub-mask using a direct selection based on the line parameters does not significantly reduce the effect of activity jitter on RV measurements. A comparison between RV RMS and precision indeed confirms that the dominant source of RV variation in the data sets is the activity jitter, which is unchanged and one order of magnitude larger than photon noise regardless of the line selection (see Table~\ref{tab:parametric}). The presence of RV shifts when using different sub-masks emphasises the importance of always using the same mask for a RV time series, i.e. always excluding lines that for some observations are blended with telluric lines or fall outside the echelle order. Finally, this analysis confirms the robustness of B$_l$ because the measurements show a clear rotational modulation and are all reasonably consistent, provided that the mask contains deep lines.


%
\section{Randomised selection}\label{sec:randomised}

\subsection{Basic principle}
For the second selection approach, we apply a randomised algorithm to extract the sub-mask minimising the RV dispersion, which is inspired by the line-by-line method developed in \citet{Dumusque2018}. We cannot follow the same exact prescription because a RV measurement on individual lines implies an increase in photon noise, which becomes relevant in the optical especially for M dwarfs. At the same time, the high density of lines in the blue part of the spectrum makes their identification more challenging.

The idea is to build sub-masks with a randomly sampled number of lines ($n_\mathrm{sample}$) from the full mask, apply LSD and derive the RV time series. Because the number of line combinations (i.e. sub-masks) is very large, we stop the iteration when each line in the full mask has been drawn at least a certain amount of times ($n_\mathrm{stop}$). An optimised exploration of the sub-mask space is performed using $n_\mathrm{sample}$ = 50 and $n_\mathrm{stop}$ = 100 (see Appendix~\ref{sec:appA} for more details), since $\sim$9000 sub-masks are examined on average and a lower $n_\mathrm{sample}$ would inevitably deteriorate the performance (Fig.\ref{fig:grid_tests}).

\begin{figure}[t]
    \centering
    \includegraphics[width=\columnwidth]{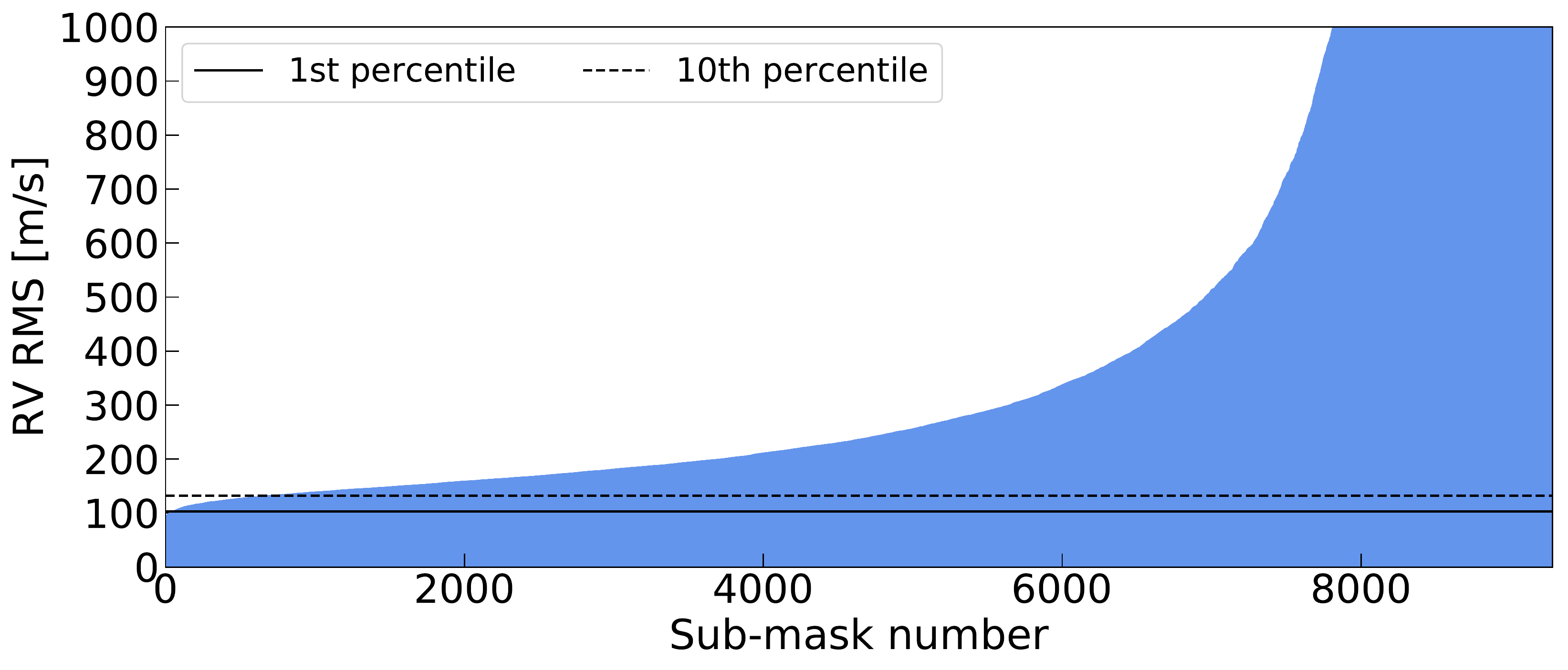}
    \caption{Distribution of the RV RMS associated with 9281 sub-masks of an example run with $n_\mathrm{sample}$ = 50 and $n_\mathrm{stop}$ = 100. Each RMS value is obtained after a 3-$\sigma$ clipping of the RV data set, to prevent outliers from contaminating the distinction between ``stable'' and ``unstable'' sub-masks. The RMS values are sorted in ascending order and truncated at 1 km s$^{-1}$ for visualisation purposes (the maximum can reach up to 100 km s$^{-1}$). Shown are the 1st (black solid line) and 10th (dashed black line) percentiles of the distribution.}
    \label{fig:randomised_sel}%
\end{figure}

The RV dispersion associated with each sub-mask is the criterion used to discriminate between ``stable'' and ``unstable'' sub-masks (Fig.~\ref{fig:randomised_sel}). From the RV RMS distribution, we isolate three groups of ``stable'' sub-masks: those with RMS lower than the 10th, 5th and 1st percentile. For each group, we find the maximum number of draws and we build a sub-mask with the lines that are drawn at least a third of this maximum number (f$_\mathrm{select}=0.33$, see Appendix~\ref{sec:appA}). Finally, we obtain the RV time series from these three sub-masks and determine the best according to the associated RMS.

Overall, two caveats stem from the randomised nature of the approach: each complete run corresponds to sampling a different region of the sub-mask (i.e. line combination) space, and it is not possible to predict exactly which of the three percentile-selected sub-masks gives the lowest RMS. In our analysis, we train the best sub-mask on the data set from 2010, since it has the largest number of observations (20), and because the activity signal is expected to lose coherency over timescales longer than months. Moreover, we do not remove lines in telluric windows from the full mask as in Sec.~\ref{sec:parametric}, to further check the reliability of the extracted best sub-mask. The starting number of lines is then 3300. The summary of the results for the analyses in the following sections can be found in Table~\ref{tab:global_summary}.

\subsection{Training for stable lines}

\begin{figure}[t]
    \centering
    \includegraphics[width=\columnwidth]{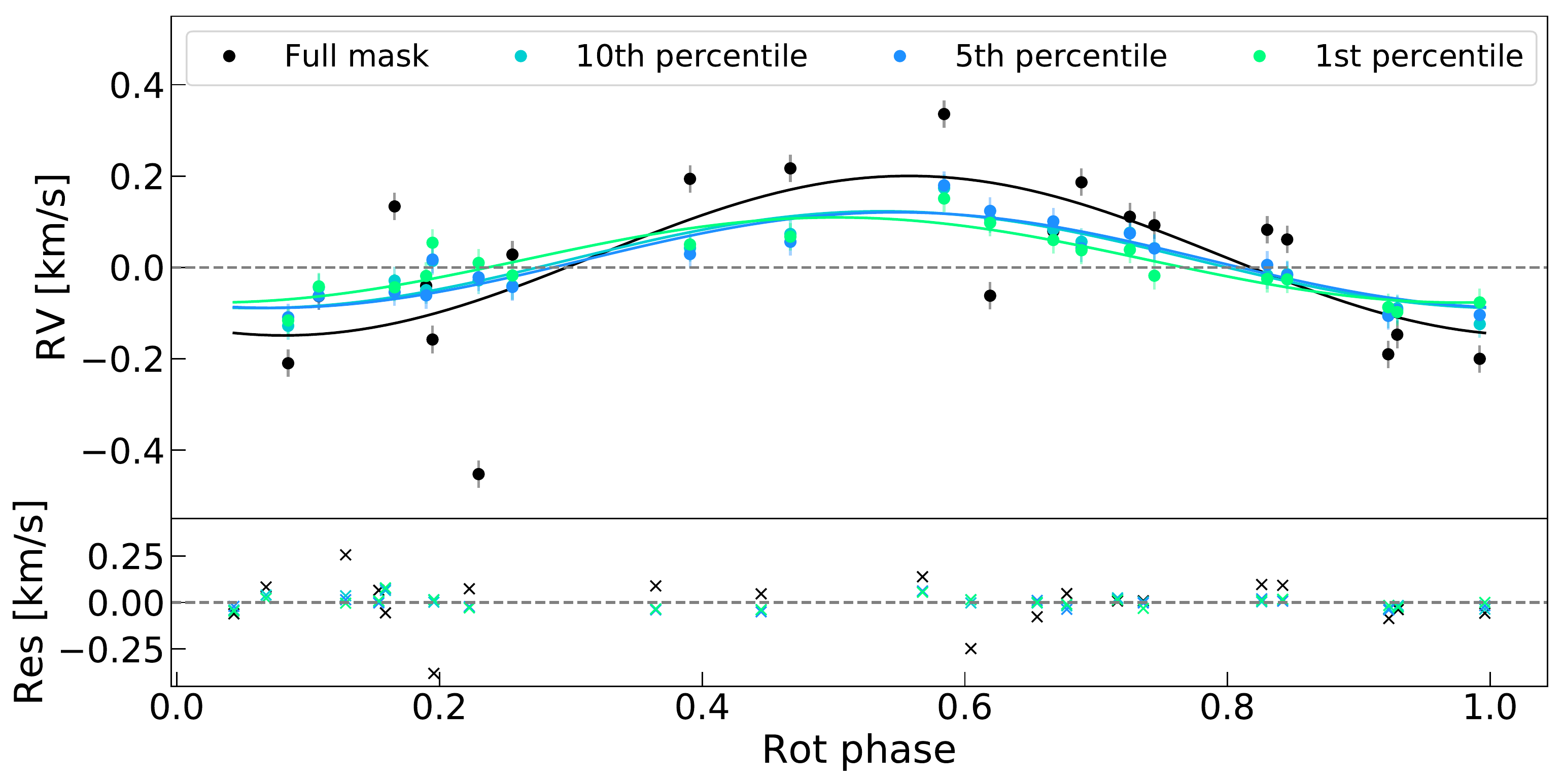}
    \caption{Output from an example run of the randomised selection. Shown are the 3-$\sigma$ clipped RV data sets computed using the full mask (black) and the sub-masks associated with the 10th (light-blue), 5th (dark blue), and 1st (green) percentiles of the RV RMS distribution.}
    \label{fig:randomised_example}%
\end{figure}

\begin{figure}[t]
    \centering
    \includegraphics[width=\columnwidth]{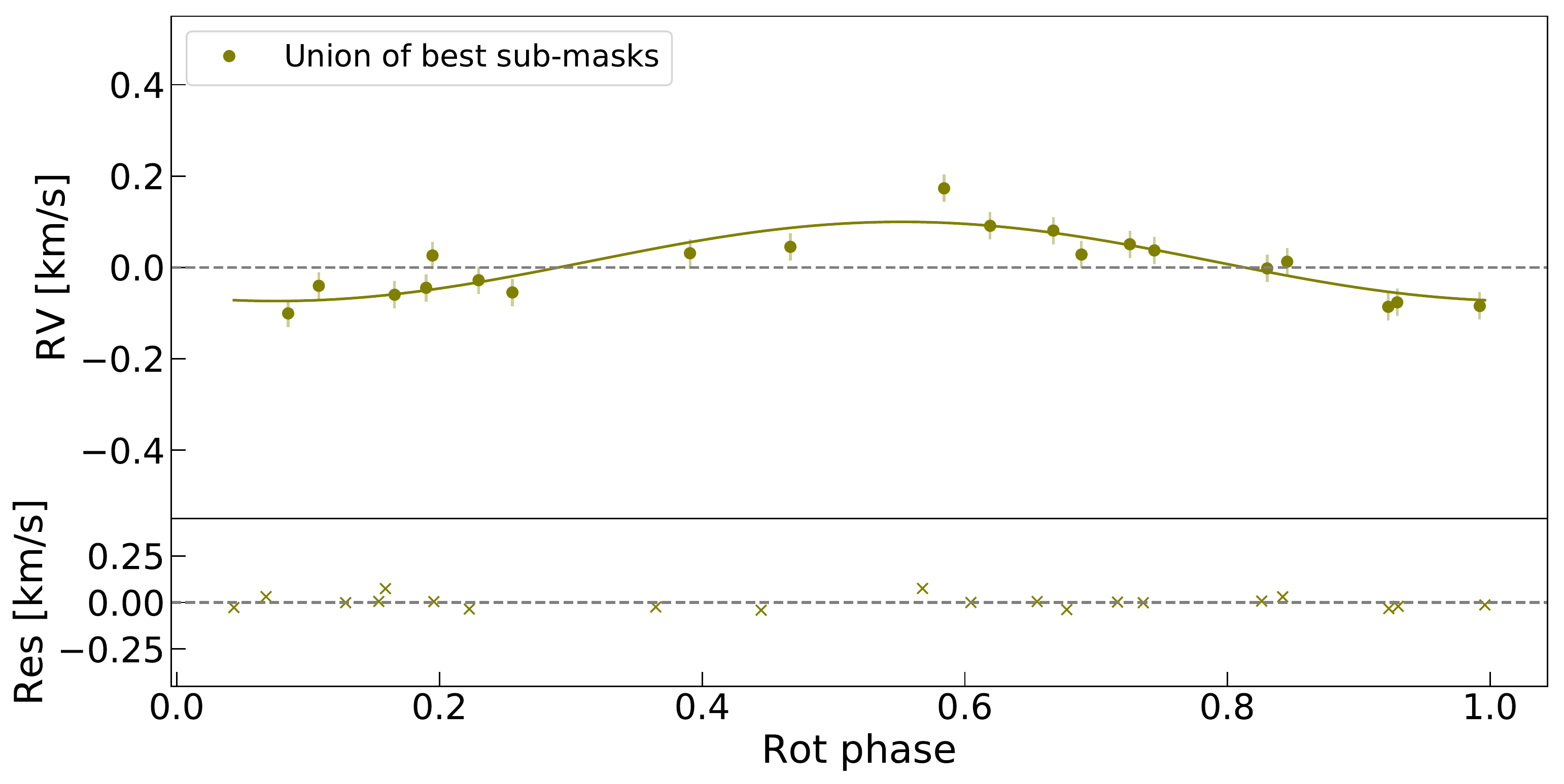}
    \caption{Phase-folded plot of the RV time series computed with the union of two best sub-masks (721 lines). The RV RMS of the data set is reduced by 63\%, and the RV semi-amplitude by 51\% relative to the full mask case. Noteworthy is also the decreased residual scatter around the fit.}
    \label{fig:randomised_RVphase}%
\end{figure}

Fig.~\ref{fig:randomised_example} illustrates the output for an example run. The full mask is associated with a RV RMS of 182 m s$^{-1}$ and a semi-amplitude of 175 m s$^{-1}$ (after performing a sinusoidal fit similarly to Sec.~\ref{sec:parametric}). The best sub-mask in this case is composed of 198 lines, none of which falls in the telluric windows, and yields a $\sim$ 63\% reduction in the RV dispersion. Considering that a $>$50\% RMS improvement is consistently achieved for multiple runs, we inspect the union of the output best sub-masks; in this case the SNR of the computed Stokes profile would benefit from an increased number of lines. We find that uniting two sub-masks (721 lines in total) reduces the RMS of the same amount as the individual ones, and that only four (out of 60) lines within the telluric windows are present in the sub-mask. Furthermore, the semi-amplitude is decreased by 51\%, as shown in Fig.~\ref{fig:randomised_RVphase}, reducing the RV modulation occurring at the stellar rotation period, and the RV precision remains around 5 m s$^{-1}$ despite the lower number of lines than the full mask. A visual comparison of the depth, wavelength and g$_\mathrm{eff}$ distributions between the full mask and the best one does not reveal any particular trend (see Appendix~\ref{sec:appB}). Therefore, no line feature can be used to single out the best sub-mask directly, in agreement with the conclusions of Sec.~\ref{sec:parametric}.

The purpose of the randomised approach is to train the best sub-mask on the year with the largest number of observations, and subsequently use it to reduce the RV dispersion of time series from other epochs. We performed tests of the sub-masks portability by using 2010-trained sub-masks on the 2007 data set (detailed in Appendix~\ref{sec:appA}) and concluded that there is approximately the same benefit as when applied to the 2010 data set.

We now compare the effects of using the 2007-trained best sub-mask on the 2007 data set with respect to the 2010-trained one. Starting from an initial 225 m s$^{-1}$ dispersion, the two cases yield a 64\% and 48\% reduction, respectively, indicating that the benefit of the randomised approach is maximised when the data set examined coincides with the training data set. Even so, applying the 2010-trained best sub-mask on 2007 leads to a substantial improvement.

When the training is applied to a less-than-ideal data set, i.e. with few and sparse observations, the portability of our approach is compromised. We indeed tested the randomised algorithm on the 2006 observations (seven in total), and obtained a sub-mask yielding an impressive 73\% improvement of the dispersion (from 249 m s$^{-1}$ to 67 m s$^{-1}$). However, when the same sub-mask was applied to the 2010 data set, we observed a severe (42\%) degradation of the RV RMS. At the same time, using this sub-mask loses meaning, as it contained only seven lines. The photon noise in this case is increased by six times relative to the full mask.

Finally, when all epochs were considered, the application of the 2010-trained best sub-mask led to a consistent RMS improvement of 45\% (from 242 m s$^{-1}$ to 134 m s$^{-1}$).

Overall, because the RMS is blind to the source of dispersion, its systematic improvement does not imply that the ``stable'' lines extracted by our randomised approach are restricted to those insensitive to activity. Indeed, telluric windows, instrumental errors (e.g. bad spectral orders), possible errors in the line list, or specific combinations of lines all introduce scatter to some degree, and are thus filtered out with our algorithm. This is noticeable when all observations are considered: in this case the activity signal has lost coherency thus, if the algorithm removes activity-sensitive lines only, using our sub-mask would be less beneficial. 



\begin{figure}[t]
    \centering
    \includegraphics[width=\columnwidth]{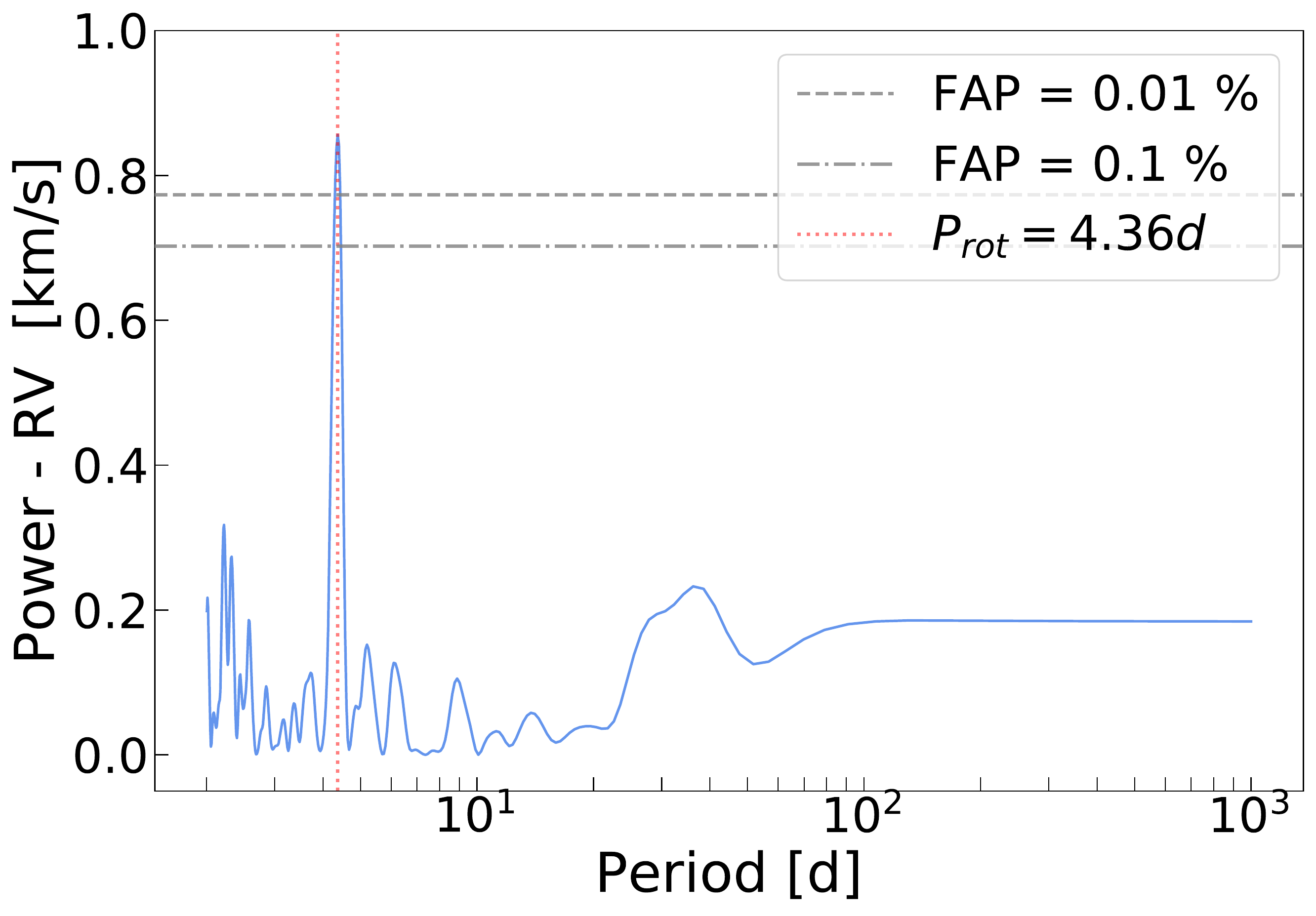}
    \includegraphics[width=\columnwidth]{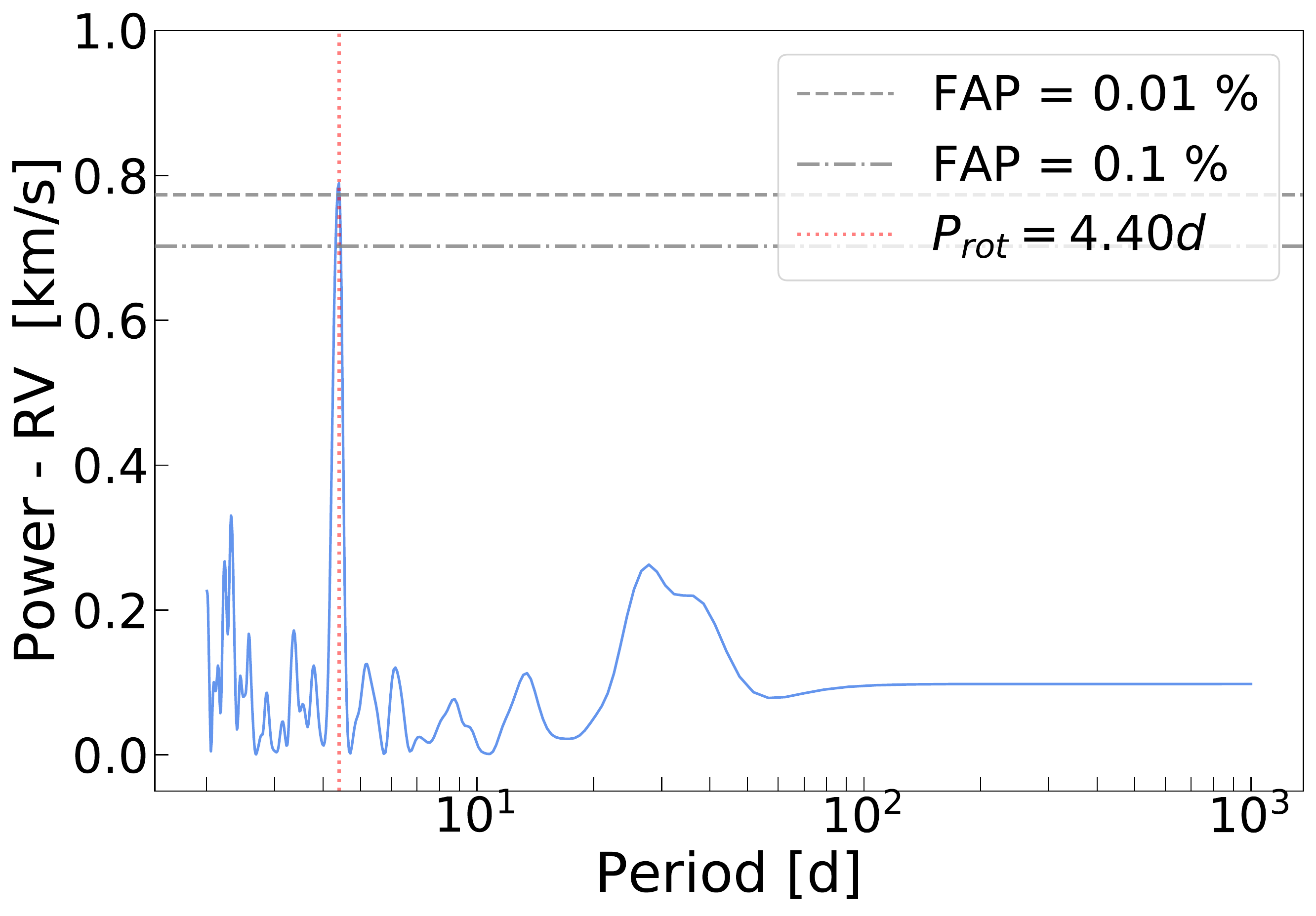}
    \caption{Top: GLS periodogram of the EV Lac 2010 RV data set obtained with the full mask. Bottom: Same but using the 2010-trained best sub-mask. We notice how the significance of the peak at the stellar rotation period decreases in the latter case, confirming the mitigating benefit of using the sub-mask derived from the randomised selection. There is a peak at 30 d showing an increase in power when the best sub-mask is applied. A period analysis of the window function \citep{VanderPlas2018} reveals that the peak is most likely due to the observing span ($\sim$2 months) and cadence of the 2010 data set.}
    \label{fig:LSPsubmask}%
\end{figure}

An additional way to show the mitigating effect of the 2010-trained best sub-mask is to inspect its impact on the GLS periodogram of the RV data sets, as illustrated in Fig.~\ref{fig:LSPsubmask}. We notice that the power (i.e. significance) of the peak associated with the stellar rotation period decreases of $\sim0.1$, implying that the corresponding activity signal is reduced. This feature occurs systematically also in the 2007 and full data sets when the 2010-trained best sub-mask is employed.

\subsection{Training for unstable lines}

We implement the randomised algorithm to work in the opposite direction, i.e. to isolate spectral lines that are susceptible to dispersive sources. If activity-sensitive lines are selected, the output data set could feature an enhanced modulation of the activity signal, therefore providing a more precise measurement of the stellar rotation period (useful for stellar physics and gyrochronology) and the possibility to model and filter the jitter efficiently.

The difference with respect to the previous training is that the sub-masks are generated considering the 90th, 95th and 99th percentile of the RV RMS distribution. Because the sub-mask exploration given by the choice ($n_\mathrm{sample}$, $n_\mathrm{stop}$)=(50,100) leads also to line combinations with extreme RV RMS values (i.e. on $>$km s$^{-1}$ level, see Fig.\ref{fig:randomised_sel}), the three sub-masks associated with these percentiles will inevitably contain a low number of lines ($<$ 10) if built using f$_\mathrm{select}=0.33$ as for the ``stable'' lines. Therefore, we require the lines to be drawn at least a tenth of the maximum number of draws for the percentile (f$_\mathrm{select}=0.1$). Finally, from each of the three sub-masks, we remove the lines that are shared with the sub-masks obtained from the 10th, 5th and 1st percentiles, respectively, to additionally deteriorate the resulting RV dispersion. 

\begin{figure}[t]
    \centering
    \includegraphics[width=\columnwidth]{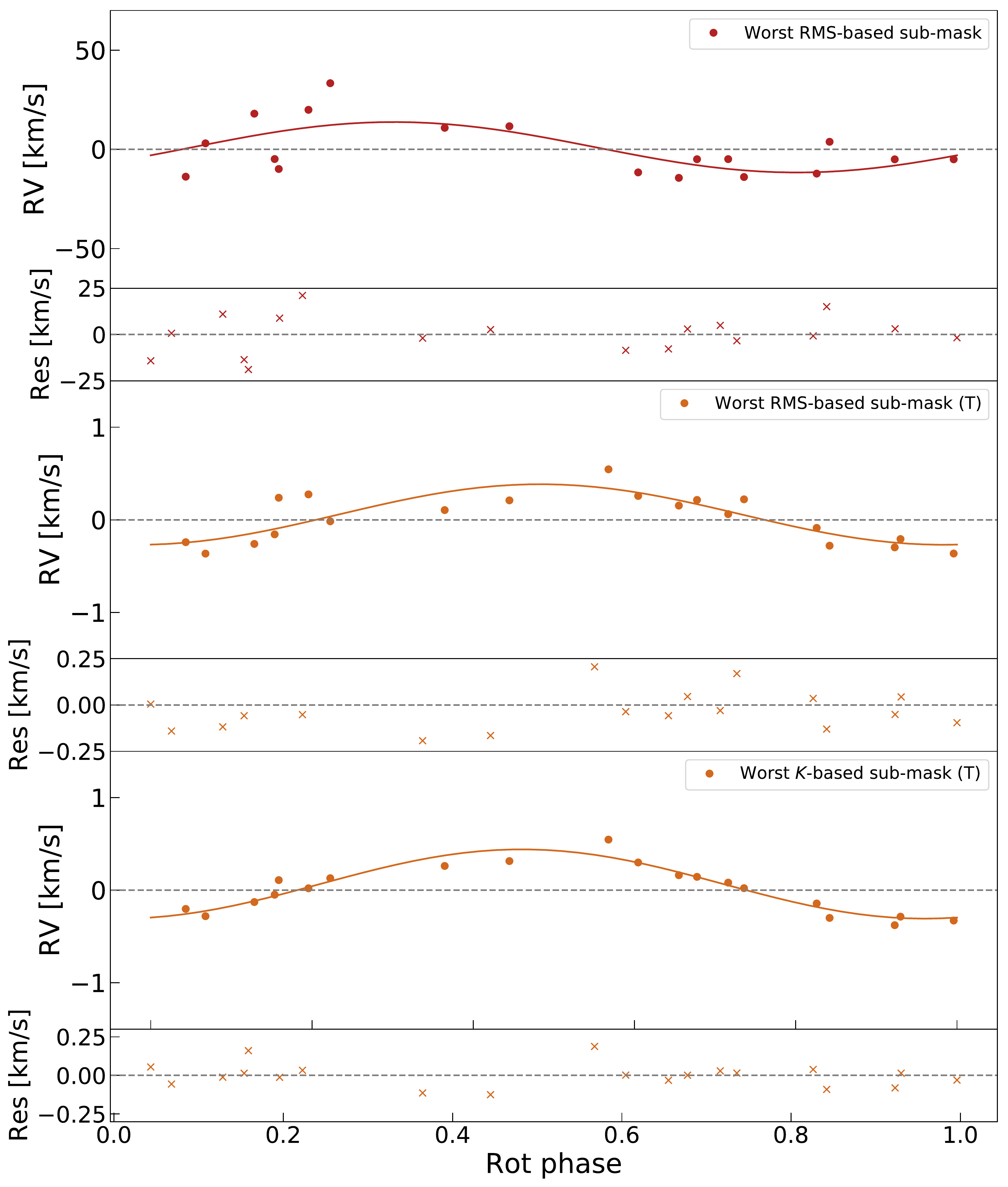}
    \caption{Phase-folded plot of the 2010 RV time series of EV Lac computed with: the worst RMS-based sub-mask including telluric windows (top), the worst RMS-based sub-mask discarding telluric windows (middle), and the worst $K$-based sub-mask discarding telluric windows (bottom). The symbol ``T'' indicates that the telluric windows are removed from the start. The first case demonstrates how lines falling in telluric windows severely contaminate the RV measurements and thus prevent a precise selection of the worst sub-mask. The second case illustrates the improvement in finding the worst sub-mask containing activity-sensitive lines, providing a clear rotational modulation of the data set. The third case shows how selecting the sub-masks based on the RV semi-amplitude, i.e. a feature that is directly connected to activity, yields the most precise selection of activity-sensitive lines. The sinusoidal fit in each panel is computed analogously to Fig.~\ref{fig:randomised_RVphase}.}
    \label{fig:randomised_badtellu}%
\end{figure}

The result is illustrated in Fig.~\ref{fig:randomised_badtellu}. The worst sub-mask has 83 lines and the associated RV dispersion is increased from 182 to 13342 m s$^{-1}$ (i.e. by 73 times). The measurements do not manifest an evident rotational modulation and span extreme values between -40 and 9 km s$^{-1}$, resulting in an artificially high RV semi-amplitude. Likewise, the union of the worst sub-masks from different runs of the algorithm does not yield useful insights.

We repeat the test, but cleaning the full mask of lines falling within telluric windows, similarly to Sec.~\ref{sec:parametric}. In this case, the algorithm produces a sub-mask of 2117 lines and with a corresponding RMS of 255 m s$^{-1}$ (Fig.~\ref{fig:randomised_badtellu}), 1.9 times larger than the initial value without telluric windows (133 m s$^{-1}$). The photon noise level is nine times lower than the RV RMS and seven times larger than when using the full mask, which could be due to a poorer modelling of the Stokes $I$ profile when distorted and ``unstable'' lines are searched. Compared to the full mask, no evident trend is displayed in the depth, wavelength and g$_\mathrm{eff}$ distributions (see Appendix~\ref{sec:appB}). Contrary to when the telluric windows are present, the RV data set shows a clearer rotational modulation, meaning that the worst sub-mask may contain more activity-sensitive lines. This marks the severe additional contamination that lines in telluric windows (or residuals from telluric correction) introduce when coexisting with the activity jitter as dispersion sources. Indeed, in that scenario, our algorithm does not precisely identify the most activity-impacted lines.

We test whether a different sub-mask selection criterion can improve the extraction of activity-sensitive lines. Instead of using the RV RMS, which does not distinguish random noise from a coherent signal, we select based on the semi-amplitude of the sinusoidal fit of the RV data set, exploiting the modulated nature of the activity signal. In this case, the two-step algorithm performs a (Levenberg-Marquardt least squares) sinusoidal fit for the RV data set corresponding to a generated sub-mask. The output sub-mask is then based on the distribution of $K$ and should contain lines that are more sensitive to stellar activity than other dispersive sources. The output RV data set features a semi-amplitude increase from 172 to 373 m s$^{-1}$, a visible rotational modulation and a reduced scatter around the sinusoidal fit (Fig.~\ref{fig:randomised_badtellu}), and a photon noise 13 times lower than the RMS, suggesting that the algorithm most likely isolates lines that are mainly sensitive to activity. 

Lastly, shallow or low-SNR lines at the bluest and reddest boundaries of the full mask may represent a possible source of confusion, as they might bias the selection of the sub-mask and pollute the resulting RV dispersion measurement. We therefore discarded lines with effective SNR (i.e., SNR multiplied by normalised depth) lower than ten and repeated the test: no improvement was recorded for the identification of the worst sub-mask and the resulting RV dispersion increased even further. This is explained considering that the bluest lines ($\sim$400 nm) are deep, and thus less likely to be affected by velocity fields (as we discussed in Sec.~\ref{sec:param_depth}). In this sense, their presence stabilises the results as indicated by a lower RV RMS of the data set. The presence (or absence) of the reddest lines ($\sim$1000 nm) has no significant impact on the results, since their number is negligible ($<20$).

\begin{figure}[t]
    \centering
    \includegraphics[width=\columnwidth]{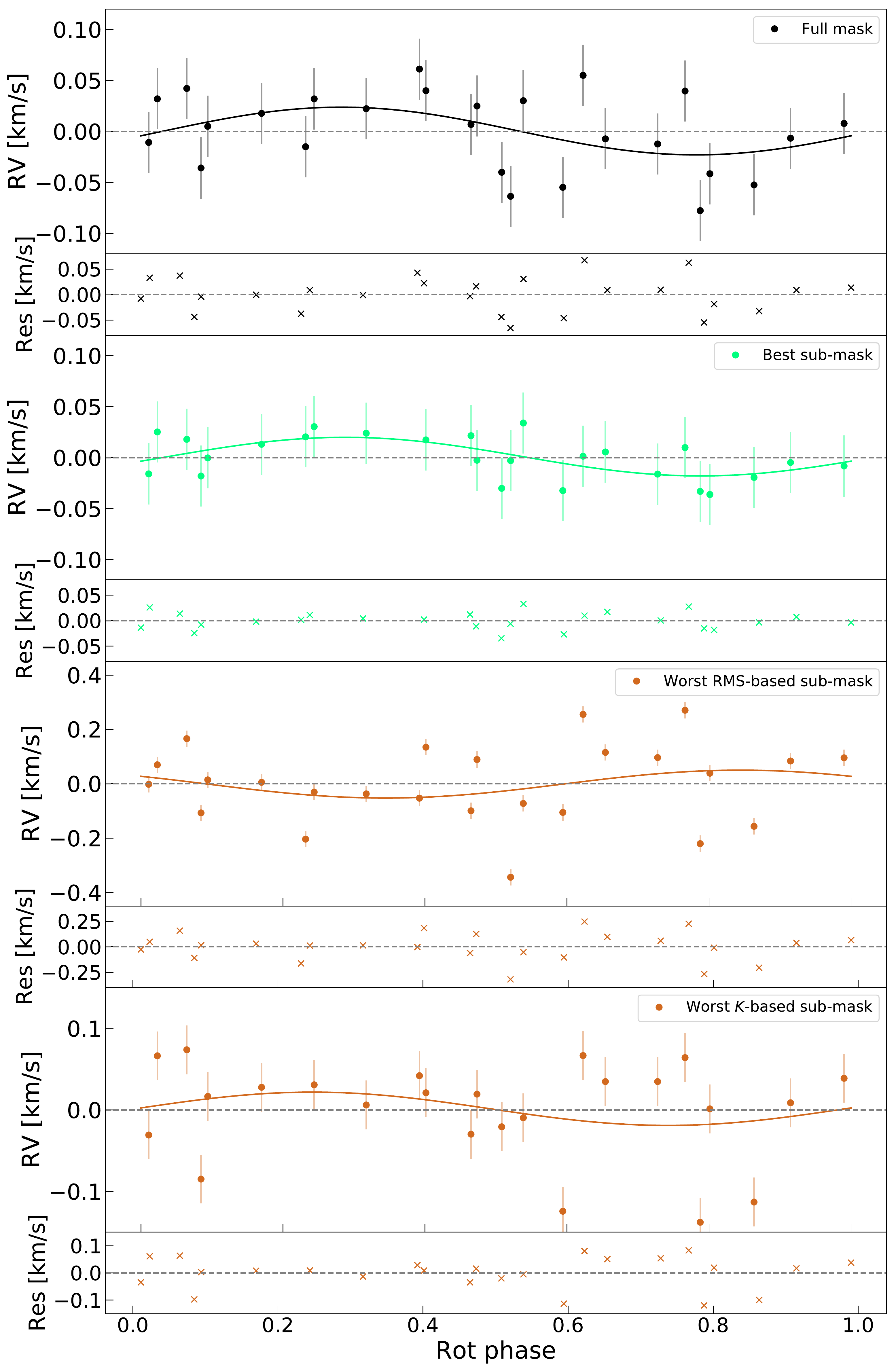}
    \caption{Phase-folded plot of the 2008 RV time series (26 observations) of DS Leo computed with (from the top): the full mask including telluric windows, the 2008-trained best sub-mask, the worst RMS-based sub-mask, and the worst $K$-based sub-mask. We use a stellar rotation period of 13.57 $\pm$ 0.04 d, computed via GLS periodogram similarly to Sec~\ref{sec:data}. From the second panel we notice a remarkable reduction of both the RV dispersion and semi-amplitude, reaching the instrumental limitation of NARVAL ($\sim$ 30 m s$^{-1}$). The third and fourth panels display an RMS-based selection which is plagued by different sources of dispersion (telluric contamination in particular), and an ineffective $K$-based selection following the low level of activity of the star, respectively.}
    \label{fig:randomised_GJ410}%
\end{figure}

\begin{table*}
\caption{Summary of all tests and simulations outlined in Sec.~\ref{sec:randomised}. The columns are: (1) the specific (sub-)mask and epoch considered, (2) the number of lines in the sub-mask, (3) the RMS of the data set, (4) the RV precision from the MC simulations, (5) the semi-amplitude of the sinusoidal fit to the data set, (6) the reduced $\chi^2$, and (7) the RMS of the fit residuals. The symbol ``T'' indicates that telluric windows are removed from the full mask. The number of observations for EV Lac are 20 in 2010, 15 in 2007, 7 in 2006, for AD Leo 14 in 2008 and DS Leo 26 in 2008.}          
\label{tab:global_summary}     
\centering                       
\begin{tabular}{l c c c c c c}      
\hline\hline                 
Case & n$_\mathrm{lines}$ & RMS & Precision & $K$ & $\chi^2_\nu$ & RMS$_\mathrm{res}$\\ 
 & & [m s$^{-1}$] & [m s$^{-1}$] & [m s$^{-1}$] & & [m s$^{-1}$]\\
\hline
\multicolumn{7}{c}{EV Lac}\\
\hline 
    Full mask on 2010                   & 3300 & 182 & 4  & 175 & 23.7 & 134\\
    Example run 10th percentile on 2010 & 666  & 79  & 11 & 105 & 1.2  & 30\\
    Example run 5th percentile on 2010  & 636  & 79  & 19 & 104 & 1.3  & 31\\
    Example run 1st percentile on 2010  & 198  & 67  & 14 & 93  & 1.2  & 30\\
    Union of best sub-masks on 2010     & 721  & 68  & 8  & 86  & 1.4  & 32\\
    Full mask on 2007                   & 3300 & 225 & 5  & 129 & 60   & 204\\
    2010-trained best sub-mask on 2007  & 198  & 118 & 17 & 142 & 6.2  & 66\\
    2007-trained best sub-mask on 2007  & 26   & 82  & 16 & 77  & 5.7  & 64\\
    Full mask on 2006                   & 3300 & 249 & 4  & 282 & 42.6 & 147\\
    2006-trained best sub-mask on 2010  & 7    & 259 & 24 & 140 & 78.4 & 244\\
    Full mask on all epochs             & 3300 & 242 & 5  & 101 & 62.7 & 231\\
    Union of 2010-trained sub-masks on all epochs    & 721 & 69 & 7 & 96 & 15.2 & 113\\
    2010-trained worst RMS-based sub-mask on 2010 & 83 & 13342 & 44 & 12718  & 1.3x$10^5$ & 10041\\
    Full mask on 2010 (T)               & 3240 & 133 & 5 & 172 & 3.3 & 50\\
    2010-trained worst RMS-based sub-mask on 2010 (T) & 2117 & 255 & 28 & 326 & 26.7 & 142\\
    2010-trained worst $K$-based sub-mask on 2010 (T) & 2209 & 246 & 19 & 373 & 7.6 & 76\\
\hline 
\multicolumn{7}{c}{AD Leo}\\
\hline 
    Full mask on 2008                  & 3300 & 110 & 2 & 80 & 14.2 & 100\\
    2008-trained best sub-mask on 2008 & 524  & 26 & 5 & 42 & 0.3 & 13 \\
    Union of EVLac-2010-trained best sub-masks on 2008 & 721 & 31 & 4 & 37 & 0.7 & 21\\
\hline 
\multicolumn{7}{c}{DS Leo}\\
\hline 
    Full mask on 2008                             & 3300 & 37  & 2 & 23 & 1.5  & 34\\
    2008-trained best sub-mask on 2008            & 571  & 20  & 3 & 18 & 0.35 & 16 \\
    2008-trained worst RMS-based sub-mask on 2008 & 800  & 143 & 6 & 51 & 24.4 & 138\\
    2008-trained worst $K$-based sub-mask on 2008 & 845  & 59  & 3 & 20 & 4.2  & 57\\
    Full mask without worst RMS-based sub-mask on 2008 & 2500 & 29 & 2 & 29 & 0.64  & 22\\
\hline                               
\end{tabular}
\end{table*}

\subsection{Training on other stars}

We examine the efficiency and consistency of the randomised approach on the active M3 dwarf AD Leo (GJ 388) and the moderately active M1 dwarf DS Leo (GJ 410). AD Leo has a reported rotation period of 2.24 days \citep{Morin2008}, i.e. approximately half of EV Lac's, and is slightly less active, with $\mathrm{L}_\mathrm{X}/\mathrm{L}_\mathrm{bol} = -3.62$ (0.5 dex less; \citealt{Wright2011}) and $\mathrm{logR'}_\mathrm{HK}=-4.00$ (0.25 dex less; \citealt{BoroSaikia2018}). AD Leo is seen almost pole-on, therefore the activity jitter is more moderate than for EV Lac. Instead, DS Leo is a slower rotator, with a period of 13.83 days \citep{Hebrard2016}, and is the least active: $\mathrm{L}_\mathrm{X}/\mathrm{L}_\mathrm{bol} = -3.80$ (0.7 dex less than EV Lac) and $\mathrm{logR'}_\mathrm{HK}=-4.16$ (0.41 dex less than EV Lac). Therefore, activity is not the dominant source of variability for DS Leo. For both AD Leo and DS Leo analysis, we do not remove the telluric windows from the initial mask.

For AD Leo, we train the best sub-mask on 14 optical observations collected in 2008. The RV dispersion of the data set computed with the full mask is 110 m s$^{-1}$, while our approach allows us to reach 26 m s$^{-1}$ (76\% improvement), i.e. the instrumental stability of NARVAL ($\sim$ 30 m s$^{-1}$, \citealt{Moutou2007}). In comparison, applying the 2010-trained best sub-mask obtained for EV Lac results in a 66\% decrease in RMS, which extends the portability of the sub-mask trained for EV Lac to other active stars, and suggests that stars with similar properties (comparable spectral type, activity level, etc.) and atmospheres have similar ``stable'' lines. When searching for the maximum benefit overall, the sub-mask training should be performed on the most dense data set of each star separately.

\begin{figure}[t!]
    \centering
    \includegraphics[width=0.95\columnwidth]{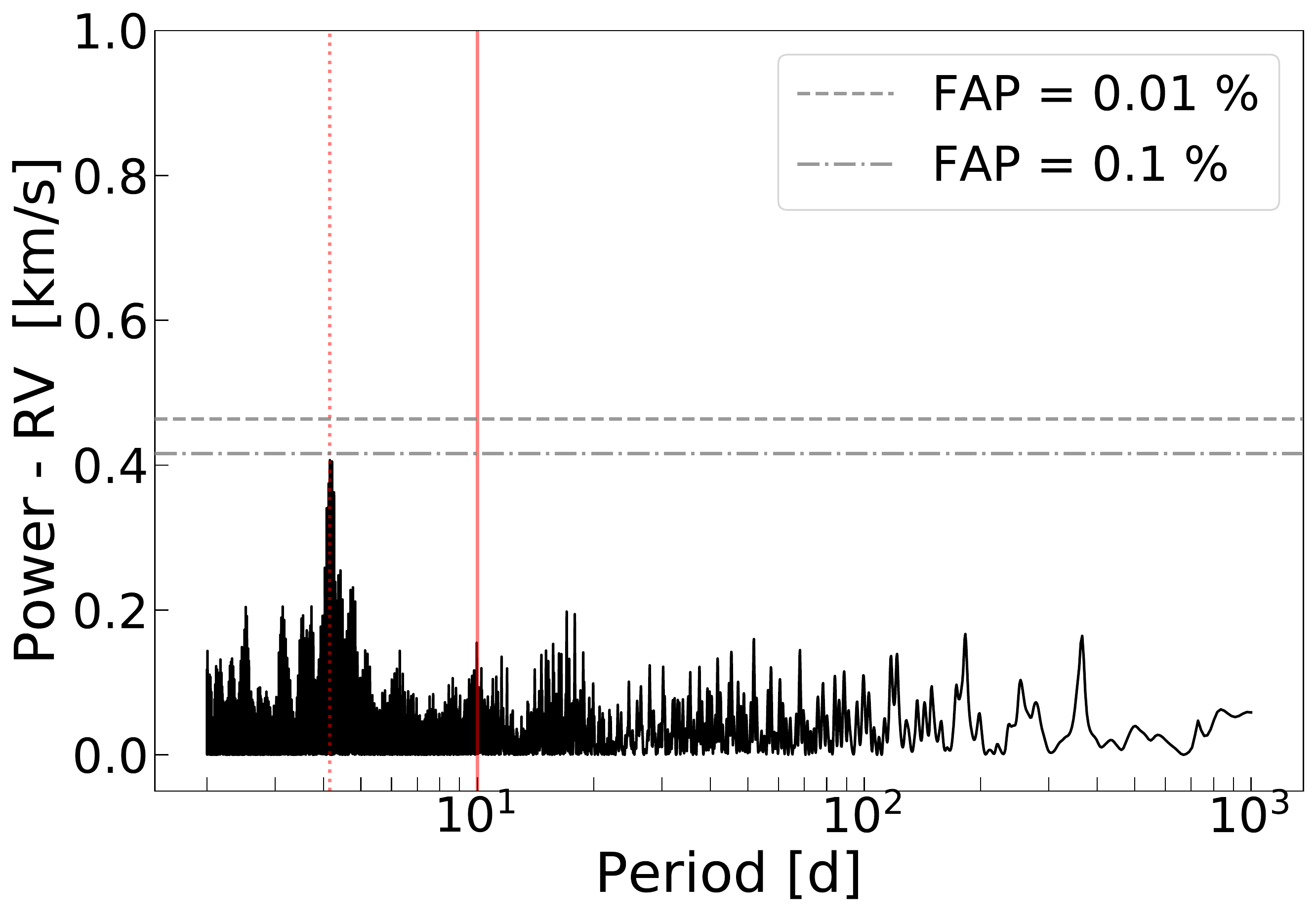}
    \includegraphics[width=0.95\columnwidth]{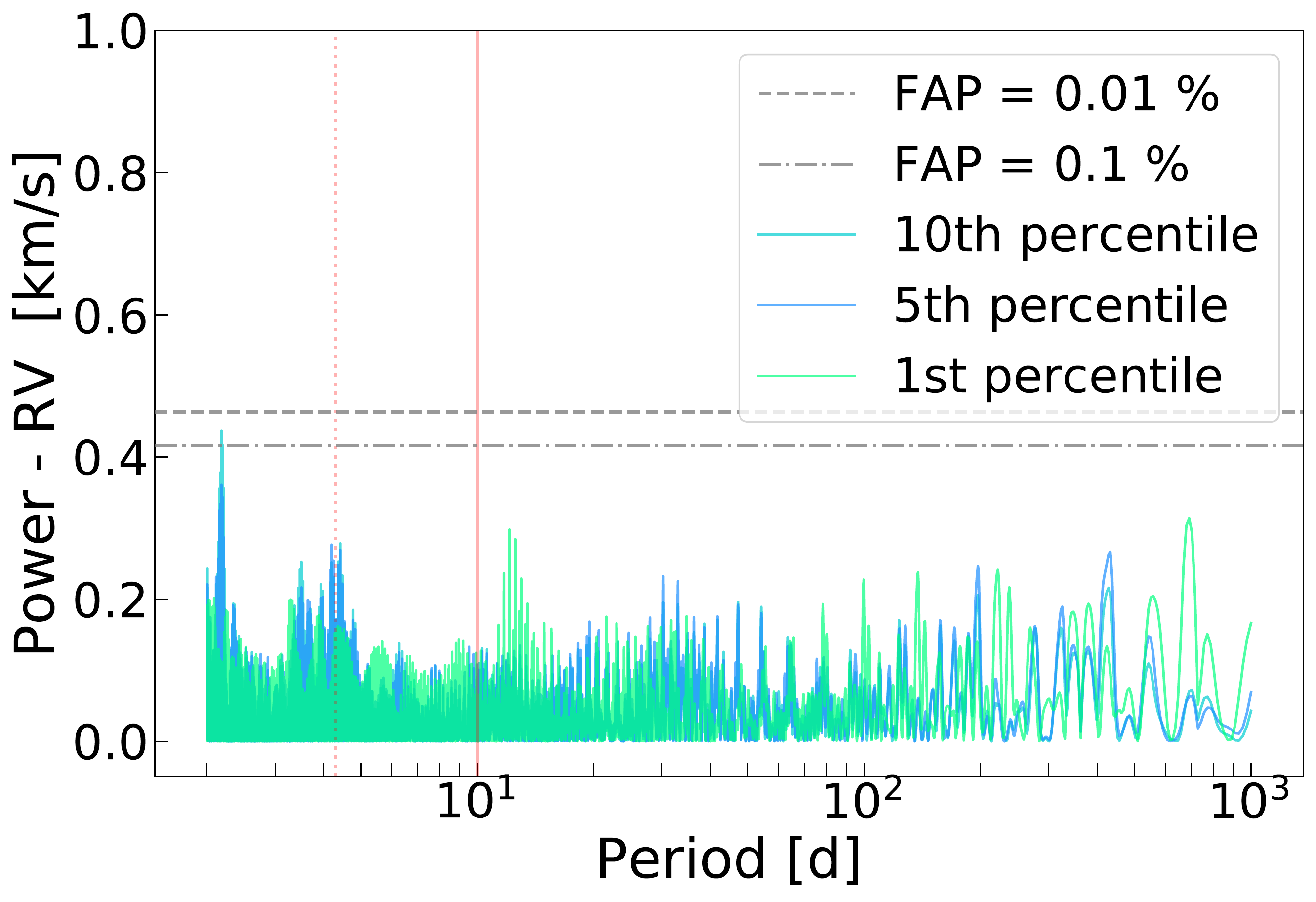}
    \includegraphics[width=0.95\columnwidth]{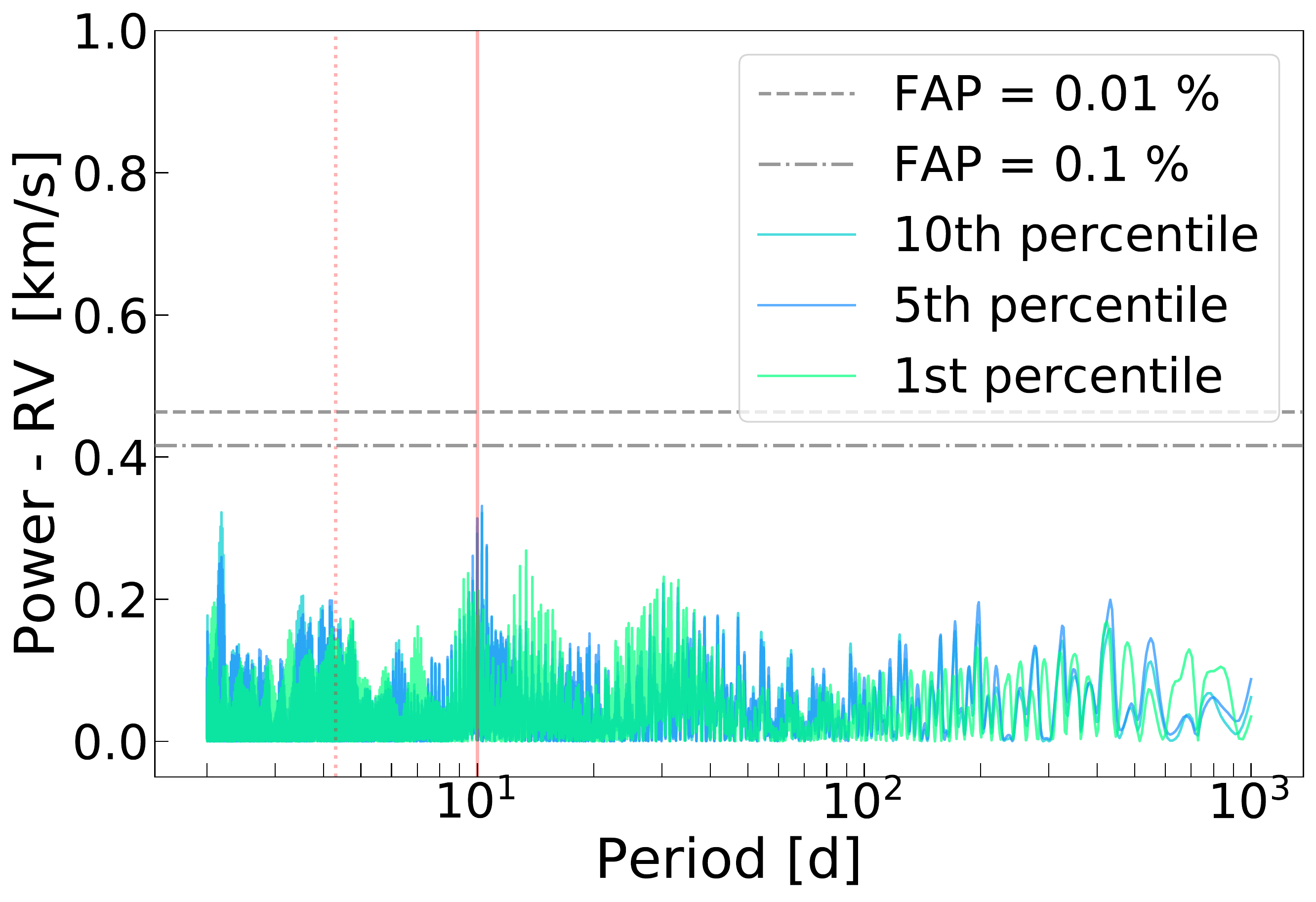}
    \caption{GLS periodograms of the full RV time series for EV Lac. Top: periodogram of the data set obtained with the full mask and a 2$\sigma_\mathrm{NARVAL}$ injected planet. Middle: periodograms of the residuals (of a sinusoidal fit at the stellar rotation period) obtained with the three 2010-trained percentile sub-masks and a 2$\sigma_\mathrm{NARVAL}$ injected planet. Bottom: periodograms of the residuals (of a sinusoidal fit at the stellar rotation period) obtained with the three 2010-trained percentile sub-masks and a 4$\sigma_\mathrm{NARVAL}$ injected planet. In each panel, the stellar rotation period and injected planetary period are indicated by a dotted and solid red line, respectively. When the 2010-trained sub-masks are used, we observe a systematic quench of the peak at the stellar rotation period and the rise of the peak at the injected orbital period as the mass of the planet increases. In the 2$\sigma_\mathrm{NARVAL}$ case, the planetary peak is not distinguishable because the dispersion is still twice larger, while in the 4$\sigma_\mathrm{NARVAL}$ case the peak becomes the highest one (with FAP$>$1\%).}
    \label{fig:evlac_gls_planets}%
\end{figure}

For DS Leo, we apply the randomised approach on 26 optical observations collected in early 2008 (Fig.~\ref{fig:randomised_GJ410}). Starting from a RV dispersion of 37 m s$^{-1}$ and semi-amplitude of 23 m s$^{-1}$ associated with the full mask, the 2008-trained best sub-mask yields a 46\% and 22\% reduction, respectively, reaching again the instrumental stability of NARVAL. This confirms that our approach is able to mitigate dispersion also on moderatively active stars, where the contribution of telluric-affected lines is more important than EV Lac and AD Leo. In comparison to \citet{Hebrard2016}, who employed HARPS-pol observations and Doppler mapping as activity-filtering technique, our values of 20 m s$^{-1}$ and 18 m s$^{-1}$ for dispersion and semi-amplitude are a factor of $2.5$ and $1.2$ higher, respectively. The possibility of combining our randomised line selection with Doppler mapping represents an appealing perspective (Bellotti et al. in prep.)

We also attempt to isolate activity-sensitive lines using both an RMS and a semi-amplitude selection criterion. As DS Leo is less active than EV Lac, the randomised algorithm should be in principle less confused by the coexistence of different sources of noise. The results of the RMS-based selection are analogous although less extreme than EV Lac, with an RV dispersion amplified by a factor of 3.9 and the absence of a clear rotational modulation. The $K$-based selection increases the RMS by 1.5 and does not affect the semi-amplitude of the fit, as expected given the lower activity level. Moreover, an inspection of the RMS-based worst sub-mask content reveals that about 20 lines fall within telluric windows, at least 10 lines more than when the randomised approach is applied on EV Lac. This further demonstrates that the randomised algorithm is capable to better discern the sources of RV noise. Finally, we remove the 800 lines of the worst RMS-based sub-mask from the full mask (using 2500 lines in total), and note a 24\% improvement in RV dispersion and a degradation of 30\% in semi-amplitude, indicating that the best sub-mask training alone is already optimal to yield a substantial improvement.

\subsection{Training with injected planets}\label{sec:planets}
We now investigate the application of the randomised algorithm when a synthetic planetary signal is included in the RV data sets. We carry out planetary recovery tests for both EV Lac and DS Leo, in order to study the detectability over two different activity regimes. 

\begin{figure}[t]
    \centering
    \includegraphics[width=\columnwidth]{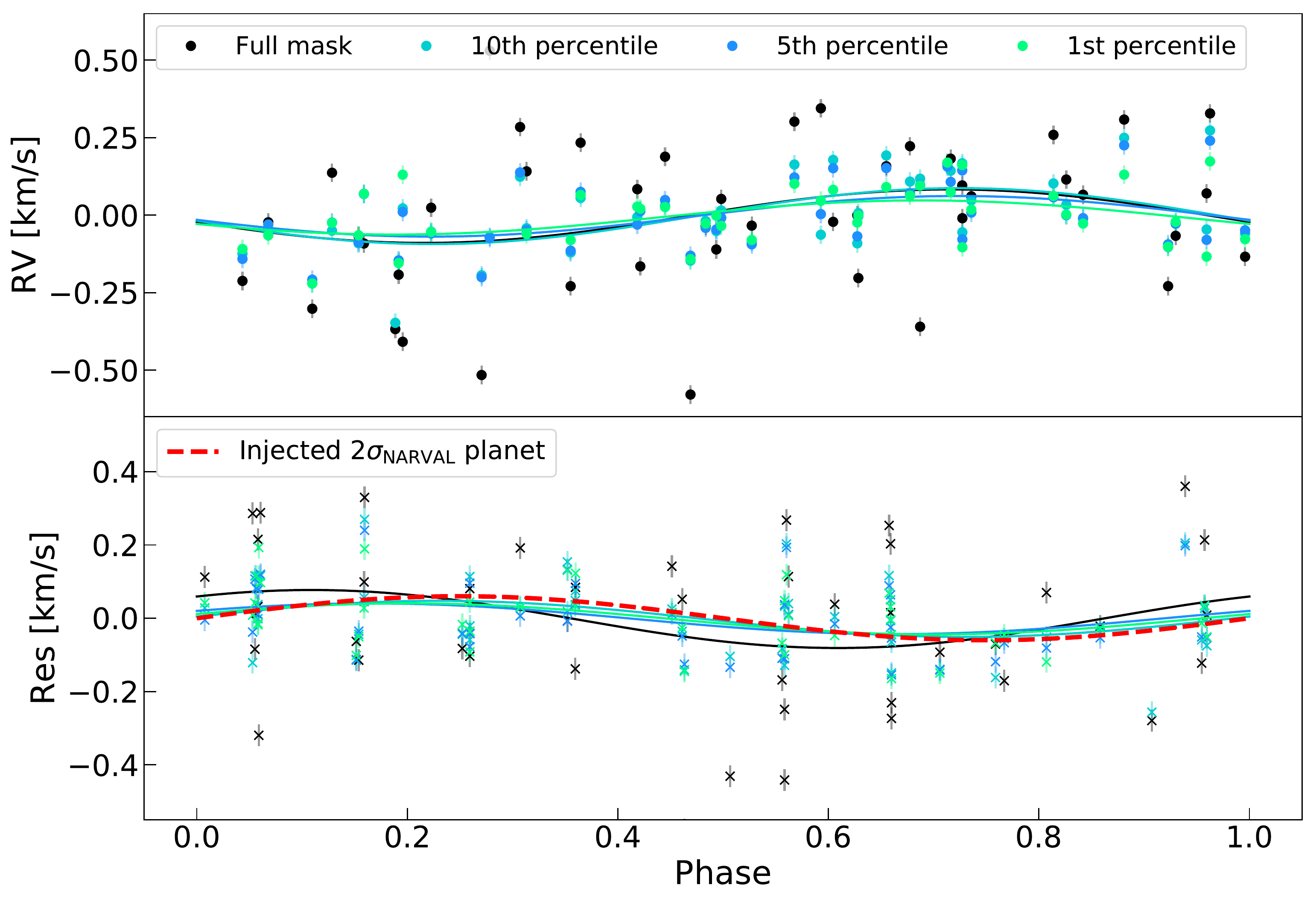}
    \caption{Phase-folded RV data sets of EV Lac when using the full mask and the 10th, 5th and 1st percentile sub-masks on all the epochs with a 2$\sigma_\mathrm{NARVAL}$ synthetic planet (red dashed line). The raw RV data sets phased at the stellar rotation period (upper panel) and the residuals to the sinusoidal fit phased at the planet orbital period (bottom panel) are illustrated. The planetary signal is evidently recovered when using the three percentile sub-masks, whereas the variability of the data set associated with the full mask does not enable a reliable recovery.}
    \label{fig:evlac_planets}%
\end{figure}

\begin{figure}[t]
    \centering
    \includegraphics[width=\columnwidth]{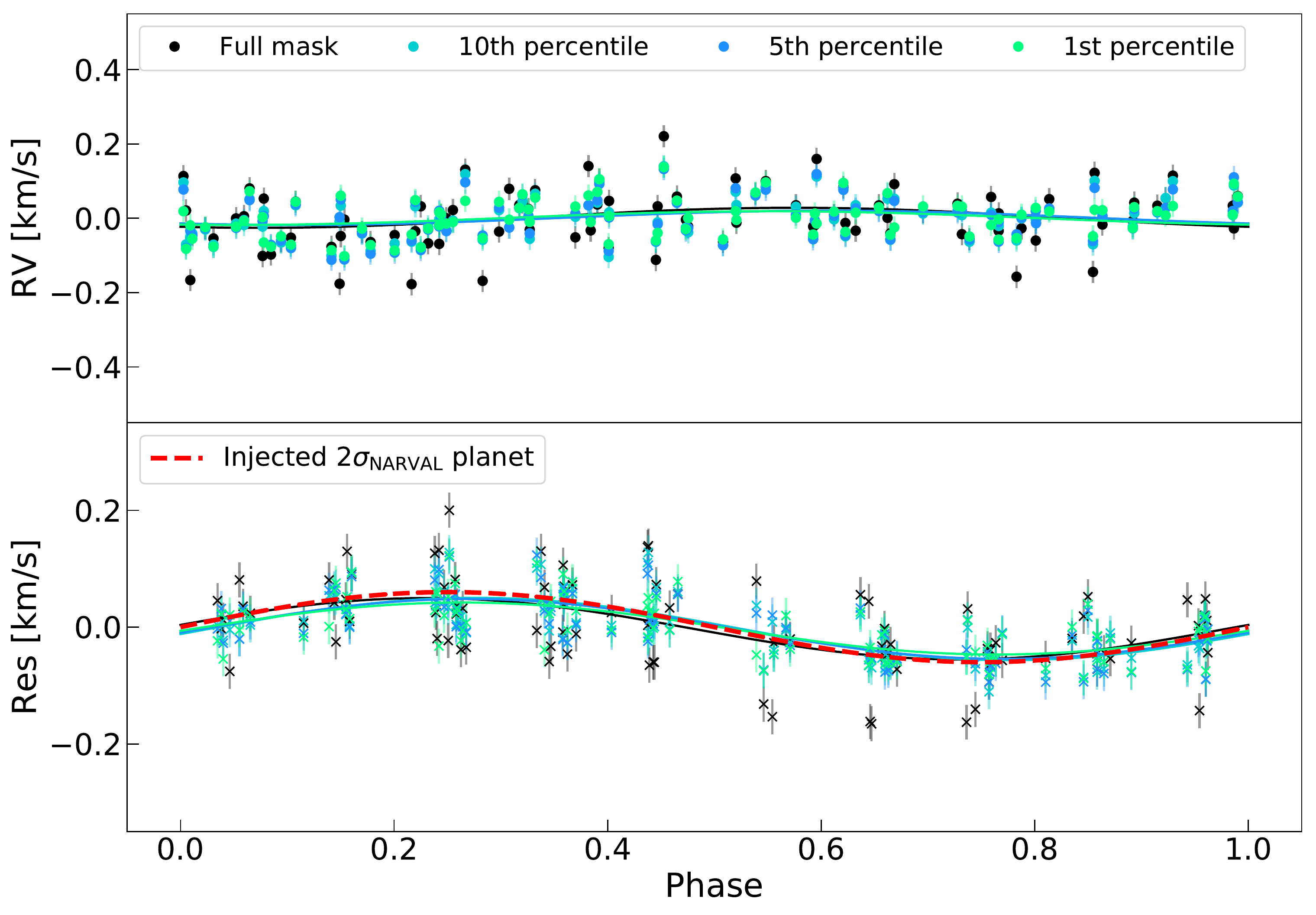}
    \caption{Same as Fig.~\ref{fig:evlac_planets} for DS Leo. We observe no substantial difference between using the full mask or the three percentile sub-masks, meaning that the planetary signal is recovered with the same confidence.}
    \label{fig:dsleo_planets}%
\end{figure}

The planet injection is applied to all the collected spectra using
\begin{equation}
\mathrm{RV}_\mathrm{p} = K_\mathrm{p}\,\mathrm{sin}\left(2\pi\left(\frac{t-T_0}{\mathrm{P}_\mathrm{orb}}\right)+\phi\right)
\label{eq:planet}    
\end{equation}
where we fixed $T_0=2450000$, $\phi$=0.0, P$_\mathrm{orb}$=10 d and we assumed circular orbits. This value of orbital period is chosen since it is different enough from the rotation period of the stars, and short enough to be clearly detected in all data sets. We considered three planets separately with a 2$\sigma_\mathrm{NARVAL}$, 3$\sigma_\mathrm{NARVAL}$, and 4$\sigma_\mathrm{NARVAL}$ signal (1$\sigma_\mathrm{NARVAL}$= 30 m s$^{-1}$, i.e. the intrinsic stability of NARVAL), and corresponding to 0.3-0.6 M$_\mathrm{Jup}$ and 0.4-0.9 M$_\mathrm{Jup}$ planets for EV Lac and DS Leo, respectively. Note that the telluric contamination is not represented realistically in these simulations, as the whole wavelength axis is coherently shifted during injection. However, this limitation is only marginal considering the negligible planet-induced shift compared to the displacement of telluric lines with respect to stellar lines throughout a year.

The training of the best sub-masks is performed on the injected spectra of the densest year for the two stars, analogously to the previous sections. The three percentile best sub-masks are then applied to the full time series comprising 57 and 93 observations spanning between 2005-2016 and 2006-2014 for EV Lac and DS Leo, respectively. Using the full time series allows us to have a time span that is more representative of a RV planet search, with an expected loss of coherency of the activity signal, but the sampling is not optimal, therefore hindering our planetary signal retrieval. The results for both stars are summarised in Table~\ref{tab:planets}; the RV precision is not reported since in all cases it is compatible with the values in Table~\ref{tab:global_summary} for each star.

\begin{table*}
\caption{Results of the planetary injection tests, for both EV Lac and DS Leo. The columns are: (1) the specific (sub-)mask considered, (2) the number of lines in the sub-mask, (3) the RMS of the data set, (4) the semi-amplitude of the sinusoidal fit to the data set (phased at the stellar rotation period), (5) the reduced $\chi^2$, and (6) the RMS of the fit residuals and (7) the semi-amplitude of the residuals (phased at the injected orbital period), (8) the reduced $\chi^2$ of the residual sinusoidal fit, and (9) the power of the injected P$_\mathrm{orb}$ peak when a GLS is applied on the residual data set. We consider a 2$\sigma_\mathrm{NARVAL}$, 3$\sigma_\mathrm{NARVAL}$, and 4$\sigma_\mathrm{NARVAL}$ circular planet on a 10 d orbit separately and train the best sub-masks on the densest epochs. The sub-masks are then used on the full time series for both stars. The symbol ``T'' indicates that the telluric windows are removed from the full mask. Formal uncertainties for the semi-amplitude are reported.}
\label{tab:planets}     
\centering                       
\begin{tabular}{l c c c c c c c c}      
\hline\hline                 
Case & n$_\mathrm{lines}$ & RMS & $K$ & $\chi^2_\nu$ & RMS$_\mathrm{res}$ & $K_\mathrm{res}$ & $\chi^2_{\nu,\mathrm{res}}$ & GLS Power\\
 & & [m s$^{-1}$] & [m s$^{-1}$] & & [m s$^{-1}$] & [m s$^{-1}$] & & \\
\hline
\multicolumn{9}{c}{EV Lac}\\
\hline                    
Full mask with 2$\sigma_\mathrm{NARVAL}$ planet       & 3300 & 244 & 96$\pm$45 & 64.7 & 234 & 90$\pm$41   & 59.2 & 0.08\\
Full mask with 2$\sigma_\mathrm{NARVAL}$ planet (T)   & 3240 & 169 & 120$\pm$27 & 24.8 & 145 & 53$\pm$28  & 23.2 & 0.07\\
10th percentile with 2$\sigma_\mathrm{NARVAL}$ planet & 1133 & 139 & 99$\pm$23 & 16.7 & 119 & 42$\pm$22   & 15.6 & 0.06\\
5th percentile with 2$\sigma_\mathrm{NARVAL}$ planet  & 385  & 131 & 88$\pm$21 & 15.4 & 114 & 44$\pm$22   & 14.3 & 0.08\\
1st percentile with 2$\sigma_\mathrm{NARVAL}$ planet  & 58   & 121 & 77$\pm$22 & 13.9 & 108 & 37$\pm$22   & 13.1 & 0.05\\
Full mask with 3$\sigma_\mathrm{NARVAL}$ planet       & 3300 & 248 & 95$\pm$46 & 66.9 & 238 & 108$\pm$41  & 59.1 & 0.12\\
Full mask with 3$\sigma_\mathrm{NARVAL}$ planet (T)   & 3240 & 174 & 117$\pm$28 & 27.0 & 151 & 82$\pm$28  & 23.1 & 0.14\\
10th percentile with 3$\sigma_\mathrm{NARVAL}$ planet & 1133 & 144 & 97$\pm$23 & 15.6 & 125 & 71$\pm$23   & 15.6 & 0.16\\
5th percentile with 3$\sigma_\mathrm{NARVAL}$ planet  & 385  & 136 & 85$\pm$22 & 17.3 & 121 & 73$\pm$23   & 14.2 & 0.18\\
1st percentile with 3$\sigma_\mathrm{NARVAL}$ planet  & 58   & 126 & 67$\pm$22 & 16.0 & 116 & 59$\pm$23   & 14.0 & 0.13\\
Full mask with 4$\sigma_\mathrm{NARVAL}$ planet       & 3300 & 253 & 92$\pm$47 & 70.1 & 244 & 131$\pm$41  & 59.0 & 0.16\\
Full mask with 4$\sigma_\mathrm{NARVAL}$ planet (T)   & 3240 & 180 & 115$\pm$30 & 30.1 & 160 & 111$\pm$28 & 23.1 & 0.23\\
10th percentile with 4$\sigma_\mathrm{NARVAL}$ planet & 1133 & 151 & 94$\pm$23 & 21.4 & 134 & 101$\pm$25  & 15.6 & 0.27\\
5th percentile with 4$\sigma_\mathrm{NARVAL}$ planet  & 385  & 144 & 82$\pm$22 & 20.2 & 131 & 102$\pm$25  & 14.2 & 0.29\\
1st percentile with 4$\sigma_\mathrm{NARVAL}$ planet  & 58   & 134 & 62$\pm$22 & 18.9 & 126 & 88$\pm$26   & 14.3 & 0.24\\
\hline
\multicolumn{9}{c}{DS Leo}\\
\hline                    
Full mask with 2$\sigma_\mathrm{NARVAL}$ planet       & 3300 & 75 & 26$\pm$11 & 6.1 & 73 & 52$\pm$9 & 4.4 & 0.27\\
Full mask with 2$\sigma_\mathrm{NARVAL}$ planet (T)   & 3240 & 57 & 20$\pm$8 & 3.6 & 55 & 58$\pm$6  & 1.5 & 0.59\\
10th percentile with 2$\sigma_\mathrm{NARVAL}$ planet & 631  & 55 & 19$\pm$8 & 3.3 & 53 & 53$\pm$5  & 1.5 & 0.55\\
5th percentile with 2$\sigma_\mathrm{NARVAL}$ planet  & 396  & 54 & 19$\pm$8 & 3.2 & 52 & 51$\pm$5  & 1.5 & 0.52\\
1st percentile with 2$\sigma_\mathrm{NARVAL}$ planet  & 566  & 50 & 18$\pm$7 & 2.7 & 48 & 44$\pm$5  & 1.4 & 0.47\\
Full mask with 3$\sigma_\mathrm{NARVAL}$ planet       & 3300 & 149 & 30$\pm$22 & 25.3 & 148 & 105$\pm$18 & 18.5 & 0.27\\
Full mask with 3$\sigma_\mathrm{NARVAL}$ planet (T)   & 3240 & 76 & 27$\pm$11 & 6.3 & 74 & 86$\pm$5  & 1.6 & 0.74\\
10th percentile with 3$\sigma_\mathrm{NARVAL}$ planet & 631  & 73 & 26$\pm$10 & 5.8 & 71 & 81$\pm$5  & 1.6 & 0.72\\
5th percentile with 3$\sigma_\mathrm{NARVAL}$ planet  & 396  & 72 & 26$\pm$10 & 5.6 & 69 & 79$\pm$5  & 1.7 & 0.71\\
1st percentile with 3$\sigma_\mathrm{NARVAL}$ planet  & 566  & 67 & 23$\pm$10 & 4.9 & 65 & 72$\pm$5  & 1.6 & 0.68\\
Full mask with 4$\sigma_\mathrm{NARVAL}$ planet       & 3300 & 106 & 41$\pm$15 & 12.1 & 102 & 108$\pm$9  & 4.8 & 0.60\\
Full mask with 4$\sigma_\mathrm{NARVAL}$ planet (T)   & 3240 & 97 & 35$\pm$14 & 10.1 & 93 & 114$\pm$6 & 1.8 & 0.82\\
10th percentile with 4$\sigma_\mathrm{NARVAL}$ planet & 631  & 93 & 34$\pm$13 & 9.3 & 90 & 109$\pm$6  & 1.8 & 0.80\\
5th percentile with 4$\sigma_\mathrm{NARVAL}$ planet  & 396  & 92 & 34$\pm$13 & 9.1 & 88 & 106$\pm$6  & 1.9 & 0.72\\
1st percentile with 4$\sigma_\mathrm{NARVAL}$ planet  & 566  & 85 & 28$\pm$13 & 8.0 & 83 & 99$\pm$6   & 1.7 & 0.79\\
\hline                                 
\end{tabular}
\end{table*}

For EV Lac, the dispersion of the RV data set computed with the full mask is 8$\sigma_\mathrm{NARVAL}$, hence no injected planetary signal is detectable. The application of the three percentile best sub-masks enables the signal to emerge instead. Compared to the full mask, for which there is more variability, the sub-masks cases feature a $>$43\% decrease in RMS, translating in a more constrained model (sinusoidal fit at the stellar rotation period) and therefore cleaner RV residuals. This is confirmed by the systematic reduction in $\chi_\nu^2$, even though its absolute estimate suggests that our method does not account entirely for the random activity. We also apply a GLS periodogram to the residuals and find that the power of the 10 d period peak increases (more significantly in the 4$\sigma_\mathrm{NARVAL}$ case) when the sub-masks are used (Fig.~\ref{fig:evlac_gls_planets}). By re-phasing the RV residuals with P$_\mathrm{orb}$, we indeed reveal the expected planetary modulation (Fig.~\ref{fig:evlac_planets}) and we extract a semi-amplitude which is lower, but consistent with the injected one within uncertainties. Moreover, we observe that the best (i.e. lowest RMS) of the three percentile sub-mask is also the one that removes the planetary signal the most (with $K$ residuals reduced by at least 30\%). This indicates a possible trade-off in the choice of the percentile sub-mask to choose, which should be considered on a case-by-case basis.

For DS Leo, the results are analogous and complementary to EV Lac, since we recover all planetary signals within error bars. With a lower activity level, the improvement represented by using the three percentile sub-masks is only marginal, and the semi-amplitude of the corresponding RV residuals is consistent with the estimate of the full mask (less than 24\% in all cases). This reinforces our finding that the randomised line downselection does not significantly suppress the planetary signal with respect to the full mask. Similarly to EV Lac, we also observe that the lowest-RMS percentile sub-mask underestimates the retrieved semi-amplitude.

Removing the telluric windows from the start has a beneficial effect for the full mask (see Table~\ref{tab:planets}), since the overall dispersion of the data set is decreased. In comparison, the output percentile sub-masks lead to both a lower dispersion and a more constrained sinusoidal fit of the RV residuals, quantified by a lower $\chi^2_\nu$ value (especially for EV Lac). We also observe similar periodogram powers at the injected P$_\mathrm{orb}$ both with the telluric-removed full mask and the three sub-masks, further indicating the our method discerns planetary signals.

We performed a similar analysis considering only the two densest epochs for each star (2007 and 2010 for EV Lac, and 2008 and 2010 for DS Leo). The idea is to exploit the increased coherence of the activity signal and obtain two separate and improved sinusoidal fits at the stellar rotation period. We combined the respective residuals into a single data set and examined its periodic content analogously to the full time series case (GLS periodogram and re-phasing at P$_\mathrm{orb}$). For EV Lac, the results with the sub-masks show no substantial change, while for DS Leo we underestimate the retrieved planetary semi-amplitude in a systematic way, probably following the low activity level of the star and thus the absence of dominant signal to filter out.

The overall positive outcome of the planetary injection tests demonstrates that our randomised approach is capable of finding the least dispersive lines without prior information on telluric windows, while preserving the planetary signal. This is especially important in view of near-infrared RV planet searches, as an accurate telluric correction in this domain is notoriously challenging \citep{Artigau2014}.


\section{Conclusions}\label{sec:conclusions}

We have carried out a study of the effects of different line masks for least-squares deconvolution on the dispersion of RV data sets, with the intent to build a mask that mitigates the effects of activity jitter. This is of particular importance for M dwarfs, given their notoriously high activity levels, and their key role in future small planet searches. Our method benefits from the multi-line nature of LSD, since it results in a high-SNR profile cleaned from line blends, and it is designed for highly dense spectra, for which the identification of individual activity-sensitive lines is more complicated.

We analyse two distinct line selection approaches for the active M dwarf EV Lac: a parametric one and a randomised one. With the parametric selection, we choose spectral lines directly based on their depth, wavelength or magnetic sensitivity, while for the randomised selection, we employ an algorithm that examines several combinations of lines and identifies those that minimise the RV dispersion. We also test whether the algorithm works in the opposite direction, i.e. to identify sub-masks containing lines affected by activity, telluric lines, and other dispersive sources. The algorithm is applied to another active star (AD Leo) and a moderatively active one (DS Leo) to validate its portability to different targets. Finally, the analysis is completed with planetary signal recovery tests for EV Lac and DS Leo. Our conclusions are summarised as follows:

\begin{enumerate}
    \item A straightforward parametric selection is not sufficient to build a sub-mask to mitigate activity, as we report no significant reduction in the RV RMS of the resulting data sets. At the same time, regardless of the parametric selections examined, the estimates of the longitudinal magnetic field are retrieved consistently, demonstrating its reliability as magnetic activity tracer and the benefit of extracting information from circularly polarised starlight. Note that the full-width at half maximum of high-g$_\mathrm{eff}$ lines can still be used to monitor the activity modulation on very active stars \citep{Klein2021}.
    \item The randomised approach allows us to generate a sub-mask with an associated 63\% decrease in RV RMS. The sub-mask is trained on 2010, i.e. the densest data set, but provides systematic RV RMS reduction when applied to 2007 (48\%) and the full time series (45\%). The added benefit is the absence of lines falling within telluric windows (without imposing a priori information) and leading to excessive dispersion, which will be particularly helpful to compensate telluric correction in the near-infrared. 
    \item The improvement of the 2010-trained best sub-mask of EV Lac can be transferred directly to another active star such as AD Leo, therefore extending the portability and possibly suggesting the definition of generally ``stable'' sub-mask. This is likely the consequence of similar properties between the stars, e.g., activity level, chemical composition and spectral type. At the same time, a remarkable and efficient reduction in RV dispersion (and semi-amplitude) is achieved when the sub-mask training is applied directly on the star of interest, demonstrating the consistency of our randomised approach.
    \item The coexistence of multiple sources of noise like high activity and telluric residuals impedes the randomised algorithm from robustly identifying ``unstable'' sub-masks containing activity-sensitive lines only, that can be potentially used to model and filter out the activity jitter. The situation improves when lines within telluric windows are removed from the start or when a RV semi-amplitude selection criterion is employed, since an evident rotational modulation at the stellar rotation period (typical of the activity signal) is present in both cases. 
    \item The planetary injection tests are successful for both EV Lac and DS Leo activity regimes. In case of a high activity level, the sub-masks built with the randomised algorithm allow to reduce the RV variability and therefore simplify the planetary signal recovery, while there is no substantial difference between using the trained sub-masks and the full one for lower activity levels. In either case, the algorithm is capable to preserve the planetary signal, with only a marginal degradation in semi-amplitude.
\end{enumerate}

The main goal of our activity-mitigating method is to increase the sensitivity toward the detection of small planets in the habitable zone of M dwarfs. In a forthcoming paper (Bellotti et al., in prep.), we will extend the analysis to the near-infrared regime and investigate the integration of the sub-mask optimal extraction with previous modelling and filtering techniques such as Doppler Imaging, Zeeman-Doppler Imaging, and Gaussian Process Regression \citep{Donati2016,Hebrard2016,Yu2017, Haywood2014}.

\begin{acknowledgements}
We thank Xavier Dumusque for detailed comments on the RV precision that enabled us to improve this paper. We thank Xavier Bonfils and Jean-Francois Donati for the insightful suggestions on the interpretation of the algorithm output.

This work is based on observations obtained at the Canada-France-Hawaii Telescope (CFHT) which is operated by the National Research Council of Canada, the Institut National des Sciences de l'Univers of the Centre National de la Recherche Scientique of France, and the University of Hawaii. This work has made use of the VALD database, operated at Uppsala University, the Institute of Astronomy RAS in Moscow, and the University of Vienna; Astropy, 12 a community-developed core Python package for Astronomy \citep{Astropy2013,Astropy2018}; NumPy \citep{VanderWalt2011}; Matplotlib: Visualization with Python \citep{Hunter2007}; SciPy \citep{Virtanen2020}. We acknowledge funding from the French National Research Agency (ANR) under contract number ANR-18-CE31-0019 (SPlaSH). XD acknoweldge for funding  in the framework of the Investissements dAvenir program (ANR-15-IDEX-02), through the funding of the ”Origin of Life” project of the Univ. Grenoble-Alpes.

\end{acknowledgements}


%
%

\bibliographystyle{aa}
\bibliography{biblio}

\begin{appendix}

\section{Control variables of the randomised approach}\label{sec:appA}

\begin{figure*}[t]
    \centering
    \includegraphics[width=\linewidth]{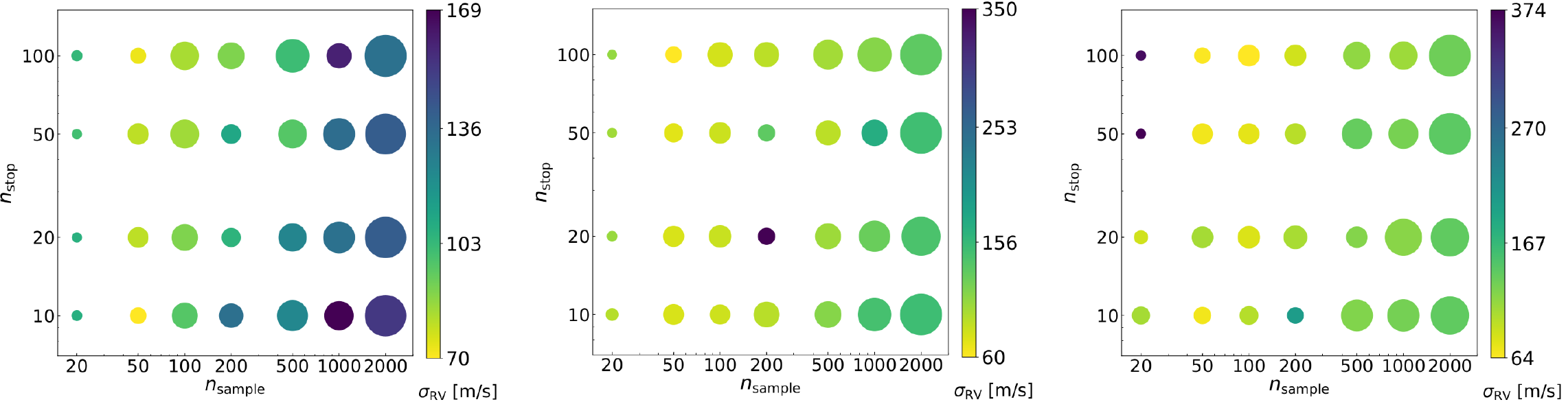}
    \caption{Optimisation of the randomised algorithm control variables. The simulation grid for each (n$_\mathrm{sample}$, n$_\mathrm{stop}$) combination is shown, with data points colour-coded and size-coded according to the RV RMS and number of lines within the sub-mask, respectively. The optimised (n$_\mathrm{sample}$, n$_\mathrm{stop}$) pair is (50, 100), ensuring both a precise RV data set and a dense exploration of the sub-mask space. Panels from left to right illustrate the 10th, 5th and 1st percentile case, respectively.
    \label{fig:grid_tests}}%
\end{figure*}

\begin{figure*}[t]
    \centering
    \includegraphics[width=\linewidth]{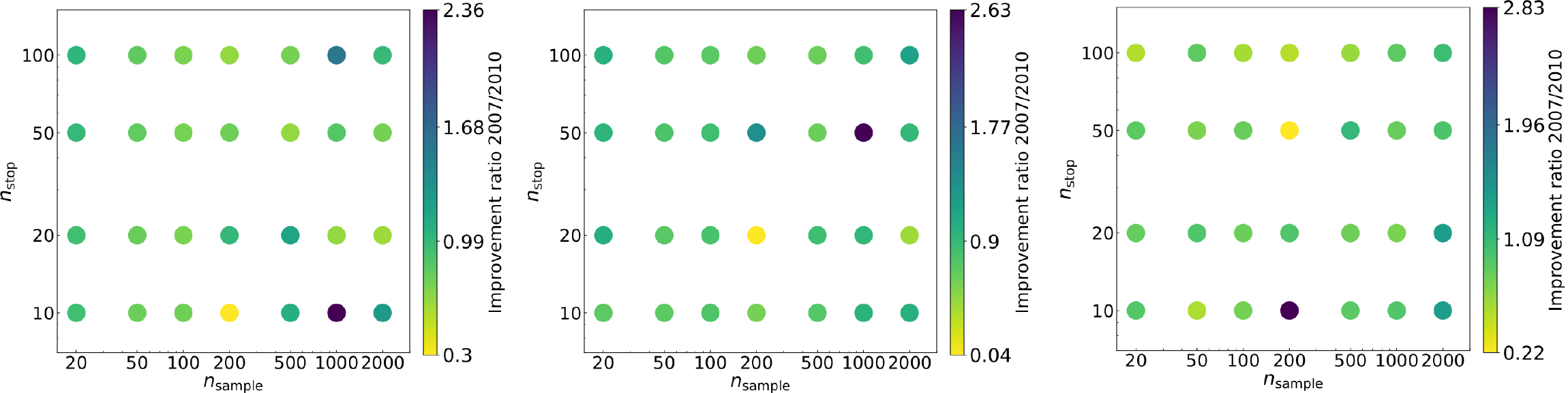}
    \caption{Testing the portability of the 2010-trained sub-masks to the 2007 data set of EV Lac, for the simulated grid of (n$_\mathrm{sample}$, n$_\mathrm{stop}$) combinations. We note that almost all the sub-masks yield the same RV RMS improvement for both 2007 and 2010 (i.e. the improvement ratio is close to 1), demonstrating the portability of the randomised approach output to other years than the training one. In few cases, the improvement ratio is either greater or lower than one, meaning that the benefit of the sub-mask is greater for 2007 or 2010, respectively. Panels from left to right illustrate the 10th, 5th and 1st percentile case, respectively. \label{fig:portability}}
\end{figure*}

In this section, we investigate the optimal control variables of the randomised approach, namely the number of sampled lines (n$_\mathrm{sample}$) and the number of times each line has to be drawn for the algorithm to stop (n$_\mathrm{stop}$). The purpose is to find a sweet-spot between the achievable reduction of RV dispersion and the number of lines in the sub-mask (n$_\mathrm{lines}$), to still obtain a moderate SNR improvement in the final LSD profile. 

We consider a grid of (n$_\mathrm{sample}$, n$_\mathrm{stop}$) combinations for the following cases: n$_\mathrm{sample}$=[20, 50, 100, 200, 500, 1000, 2000] and n$_\mathrm{stop}$=[10, 20, 50, 100]. For each value of n$_\mathrm{sample}$, we apply the randomised approach with n$_\mathrm{stop}$=100 to obtain the largest sample of sub-masks or, equivalently, to explore the largest region of line combination space. Then, because the n$_\mathrm{stop}$=100 run contains the lower n$_\mathrm{stop}$ cases by construction, we simply extract the corresponding sub-masks. This procedure enables us to save computational time and ensure consistency when comparing results given a certain n$_\mathrm{sample}$. The number of sub-masks examined for the grid are reported in Table~\ref{tab:nsubmasks_grid}.

Given that our randomised algorithm examines the 1st, 5th and 10th percentiles of the RV RMS distribution to build the final sub-masks (see Sec.~\ref{sec:randomized}), the output of the simulation grid is threefold, as illustrated in Fig.~\ref{fig:grid_tests}. In all cases, the number of lines in the sub-mask decreases systematically with a smaller number of samples (i.e. lower n$_\mathrm{sample}$), while it is not affected by n$_\mathrm{stop}$. The RV RMS correlates mainly with the number of sampled lines, reaching a minimum at n$_\mathrm{sample}$=50 and rising again for n$_\mathrm{sample}$=20, presumably due to the photon noise contribution when a small number of lines is used. The RMS also improves with increasing n$_\mathrm{stop}$, until it stabilises between n$_\mathrm{stop}$ = 50 and 100. Combining these features, we are able to locate the optimised control variables at (n$_\mathrm{sample}$, n$_\mathrm{stop}$) = (50, 100), ensuring a dense and statistically robust exploration of the sub-mask space. A visual inspection of the Stokes $I$ profiles resulting from the best sub-masks of the grid reveals that our algorithm identifies deep and sharp lines reliably when n$_\mathrm{sample}$ is decreased, which validates further our optimisation. From Fig.~\ref{fig:grid_tests} we note that the output of our randomised algorithm can be strategically adapted according to the SNR tolerance of the observations, i.e., the best sub-mask is chosen based on both the number of lines (hence, SNR) and achievable RV precision.

\subsection{Portability to other epochs}
The simulated grid of sub-masks enables us to test their portability to other epochs in which the star has been observed. In particular, we examine whether the RV RMS improvement associated with each sub-mask can be transferred to the 2007 time series (15 observations), by performing LSD on this data set with the 2010-trained sub-masks of the grid. We then compute the RMS improvement ratio between 2007 and 2010, as shown in Fig.~\ref{fig:portability}. Except for a couple of outliers, indicating sub-masks with a predominant benefit for a certain year, we observe a general ratio close to 1.0, demonstrating that the RMS improvement is the same for both years. Furthermore, we note the absence of outliers when n$_\mathrm{stop}$ is large, confirming our previous choice of n$_\mathrm{stop}$=100.

\begin{table}
\caption{Number of sub-masks examined for each (n$_\mathrm{sample}$, n$_\mathrm{stop}$) combination of the simulation grid.}
\label{tab:nsubmasks_grid}    
\centering  
\begin{tabular}{c c c c c c c c}
\hline\hline
$n_{\mathrm{sample}}$ & 20 & 50 & 100 & 200 & 500 & 1000 & 2000\\
$n_{\mathrm{stop}}$ & & & & & & & \\
\hline
	10 & 3918 & 1689 & 857 & 394 & 161 & 72 & 34\\
	20 & 6448 & 2504 & 1389 & 715 & 244 & 116 & 54\\
	50 & 13464 & 5295 & 2588 & 1342 & 521 & 233 & 108\\
	100 & 23208 & 9281 & 4549 & 2345 & 880 & 447 & 200\\
\hline
\end{tabular}
\end{table}

\subsection{Selection of the best sub-masks}
In our randomised algorithm, the ``stable'' sub-masks are isolated based on the the 10th, 5th and 1st percentiles of the RV RMS distribution. Then, the three final sub-masks (one for each percentile) are built including lines that are drawn at least a fraction (f$_\mathrm{select}$) of the maximum number of draws for the specific percentile. f$_\mathrm{select}$ is therefore an additional variable to be optimised, as it determines the number of lines in the best sub-mask and affects the achievable RV precision. 

\begin{figure}[t]
    \centering
    \includegraphics[width=0.81\linewidth]{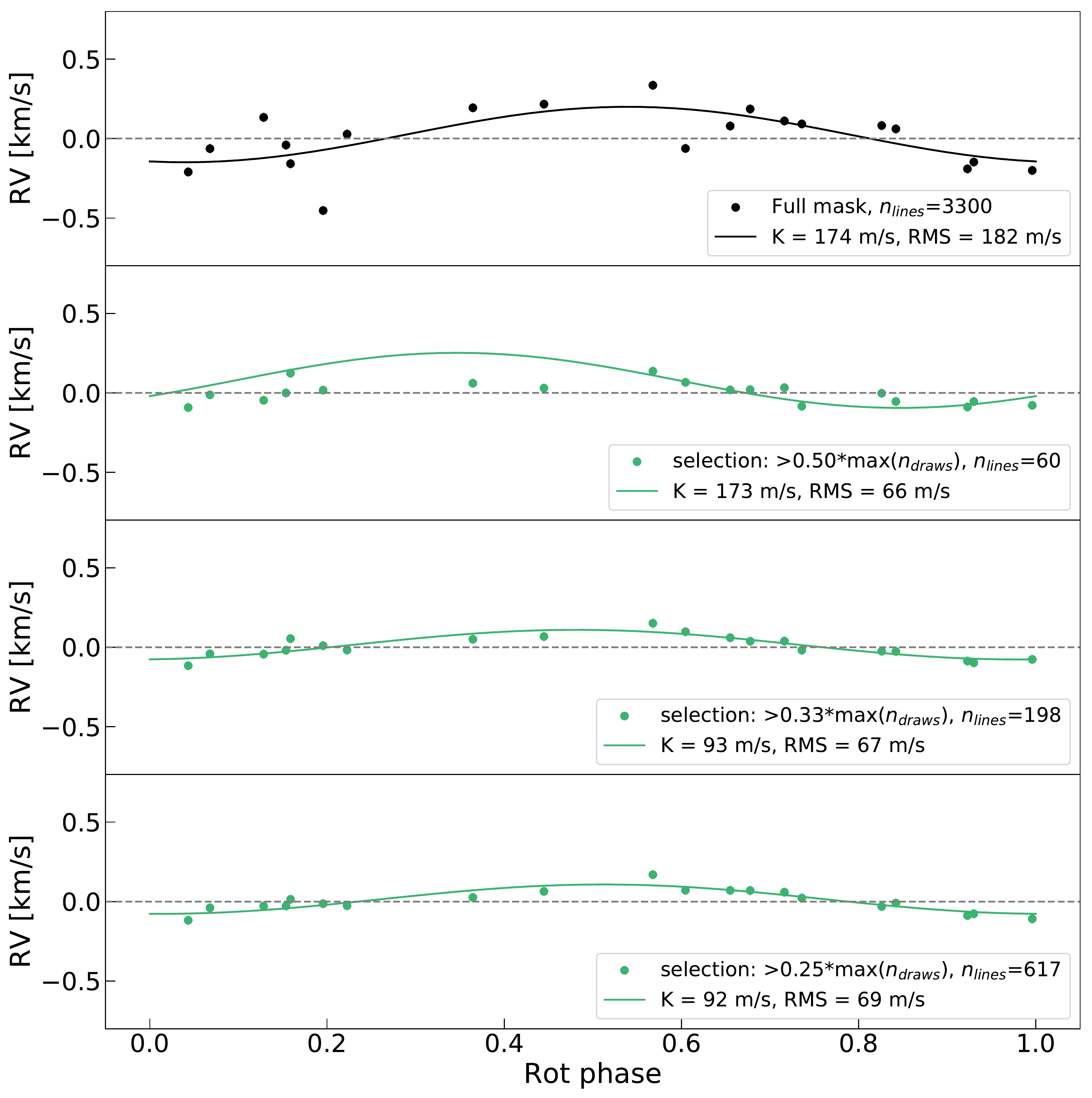}
    \caption{Test of the f$_\mathrm{select}$ variable in the randomised algorithm, i.e. the fraction determining the number of draws each line must have to be included in the three percentile sub-masks. From top to bottom: using the full mask, f$_\mathrm{select}$=0.5 sub-mask, f$_\mathrm{select}$=0.33, and f$_\mathrm{select}$=0.25. All the f$_\mathrm{select}$ cases refer to the 1st percentile sub-mask, as the other percentiles show deteriorated, but analogous features. In each panel the solid line indicates the sinusoidal fit at the period found in Sec.~\ref{sec:data}. We observe that f$_\mathrm{select}$=0.33 is the best trade-off between number of lines included and achievable RV RMS. \label{fig:fselect}}
\end{figure}

The simulation tests for (n$_\mathrm{sample}$, n$_\mathrm{stop}$) in the previous section were performed with f$_\mathrm{select}$=0.5. Here, we start from the optimised (n$_\mathrm{sample}$, n$_\mathrm{stop}$)=(50,100) run and compare the results using f$_\mathrm{select}$=0.25, 0.33, and 0.5. From Fig.~\ref{fig:fselect}, we realise that f$_\mathrm{select}$=0.5 is a strong constraint, given that we obtain a similar RV RMS for the other two cases, but with smaller number of lines. Instead, the f$_\mathrm{select}$=0.25 is a weak constraint, yielding a deterioration in both RMS and semi-amplitude; this is evident with additional runs of the randomised algorithm. We therefore conclude that f$_\mathrm{select}$=0.33 is the appropriate value to obtain a substantial improvement in RMS while maintaining a moderatively large number of lines in the final sub-masks.

Selecting lines for the three percentile sub-masks based on a fraction of the maximum number of draws (for that percentile) may lead to a lower number of lines in the e.g. 5th percentile relative to the 1st. In fact, if only few sub-masks fall below the 1st percentile of the RV RMS distribution, the maximum number of draws will inevitably be small and consequently more lines will be selected in the corresponding percentile sub-mask.

Finally, we tested an alternative method to build the three percentile sub-masks and use them in LSD. Instead of rejecting lines that do not satisfy the constraint imposed by f$_\mathrm{select}$, we kept all the lines of the sub-masks within a percentile and employed the number of draws as an additional weight in the LSD computation. More precisely, to compute the Stokes $I$ profile, each line in the mask is weighted based on depth \citep{Kochukhov2010}
\begin{equation}
    w_I = \frac{d}{d_n}
    \label{eq:weight}
\end{equation}
where $d_n$ is the normalisation depth (used to ensure $w_I=1$). We therefore updated Eq.\ref{eq:weight} by multiplying it with the number of draws. We found a systematic deterioration in both RMS and semi-amplitude for all three percentile sub-masks, relative to the f$_\mathrm{select}$=0.33 case, of about 20-30\%.

\section{Properties of the trained masks}\label{sec:appB}

\begin{figure*}[t]
    \centering
    \includegraphics[width=\linewidth]{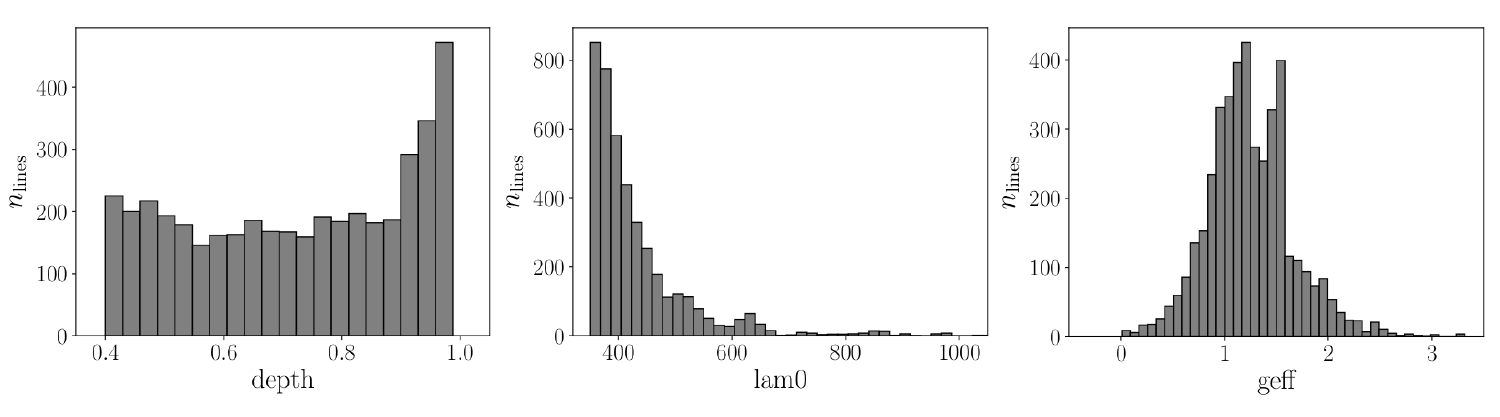}
    \includegraphics[width=\linewidth]{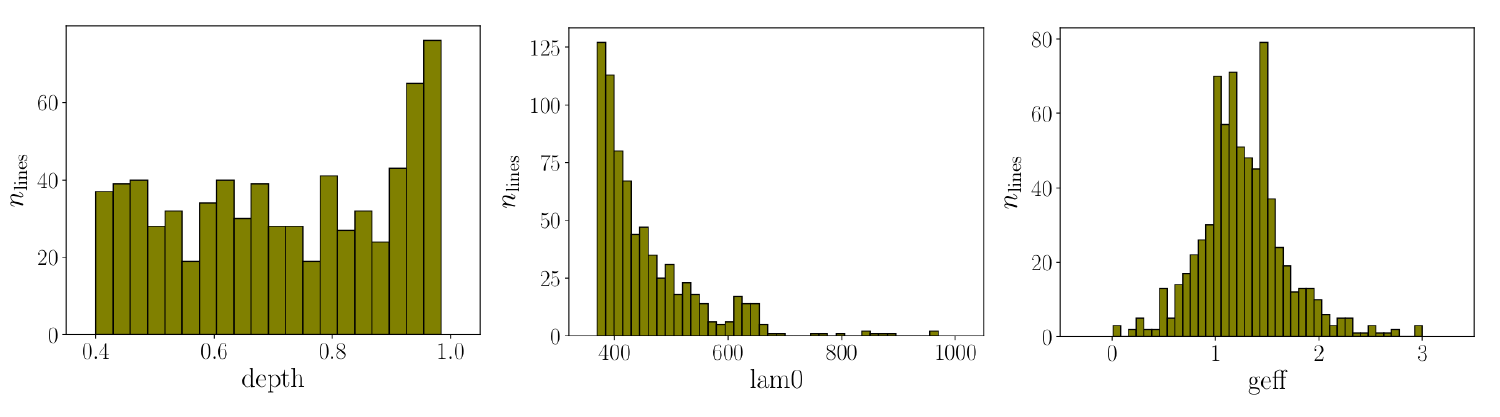}
    \includegraphics[width=\linewidth]{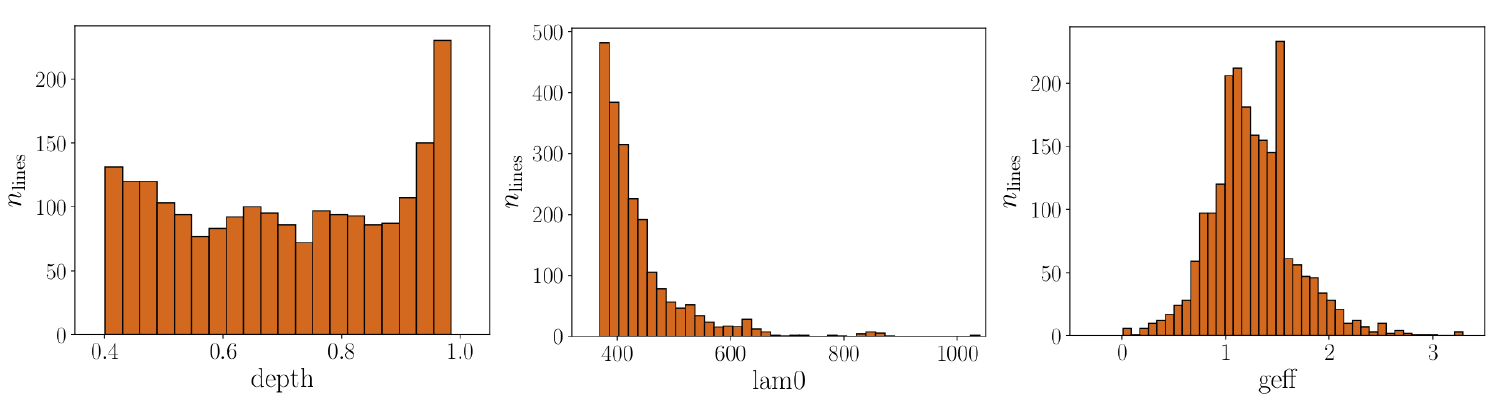}
    \caption{Distributions of the line parameters for three examined masks: the full mask without telluric windows a priori (top row), union of best sub-masks (central row), and 2010-trained worst RMS-based sub-mask without telluric windows a priori (bottom row). The parameters shown are (from left to right) depth, wavelength and Land\'e factor. The statistics (mean, median and standard deviation) of the depth distributions is ($\mu$=0.73, $\tilde\mu$=0.74, $\sigma$=0.19), ($\mu$=0.72, $\tilde\mu$=0.72, $\sigma$=0.18), and ($\mu$=0.71, $\tilde\mu$=0.71, $\sigma$=0.19) for the three masks, respectively. Analogously, the wavelength distribution statistics (in nm) is ($\mu$=433.1, $\tilde\mu$=401.2, $\sigma$=97.9), ($\mu$=454.6, $\tilde\mu$=424.3, $\sigma$=88.9), and ($\mu$=435.9, $\tilde\mu$=412.3, $\sigma$=75.8) and the g$_\mathrm{eff}$ statistics is ($\mu$=1.25, $\tilde\mu$=1.21, $\sigma$=0.41), ($\mu$=1.27, $\tilde\mu$=1.24, $\sigma$=0.42), and ($\mu$=1.26, $\tilde\mu$=1.22, $\sigma$=0.41). \label{fig:line_distribs}}
\end{figure*}

\begin{figure*}[t]
    \centering
    \includegraphics[width=0.312\linewidth]{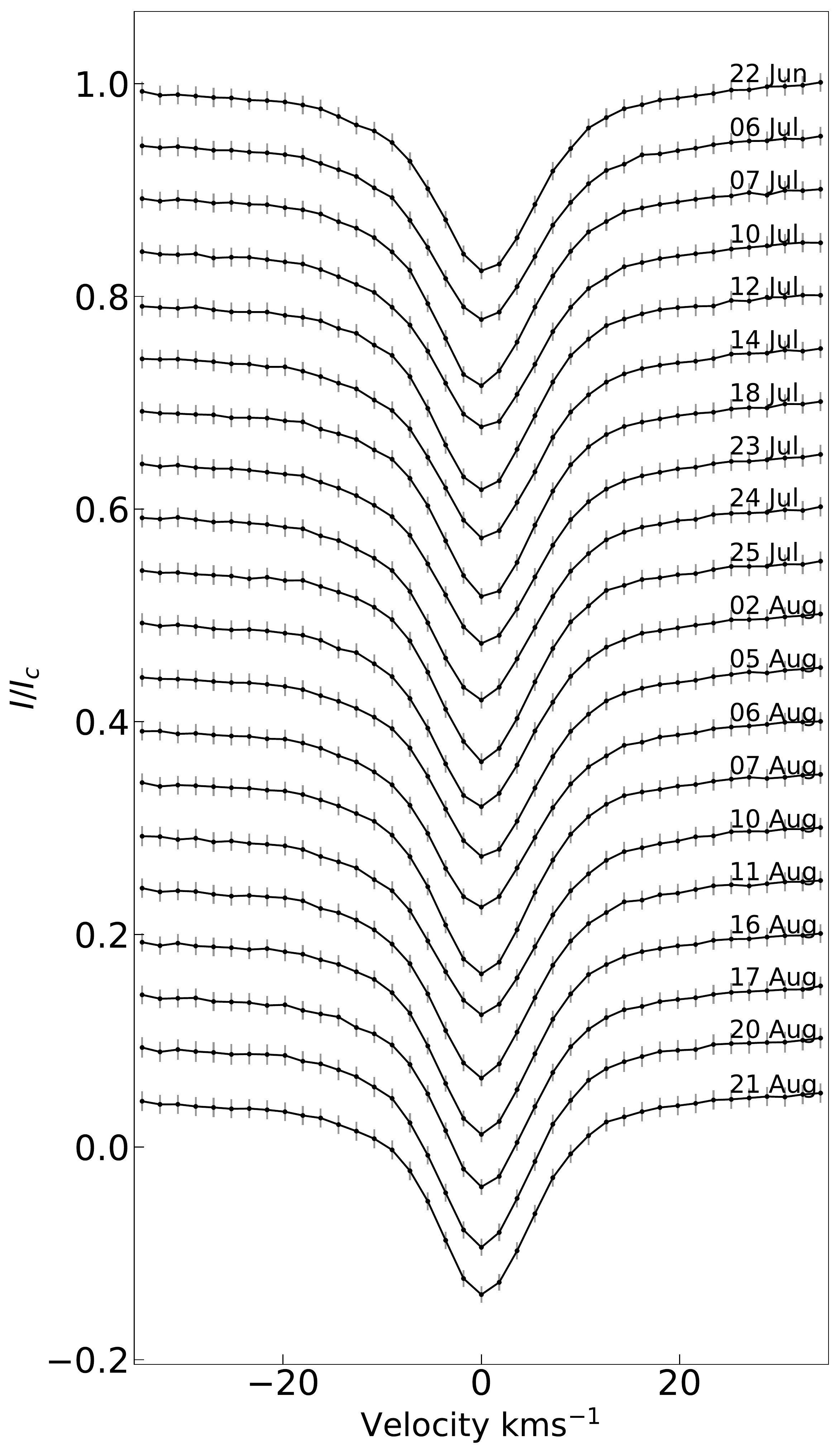}
    \includegraphics[width=0.312\linewidth]{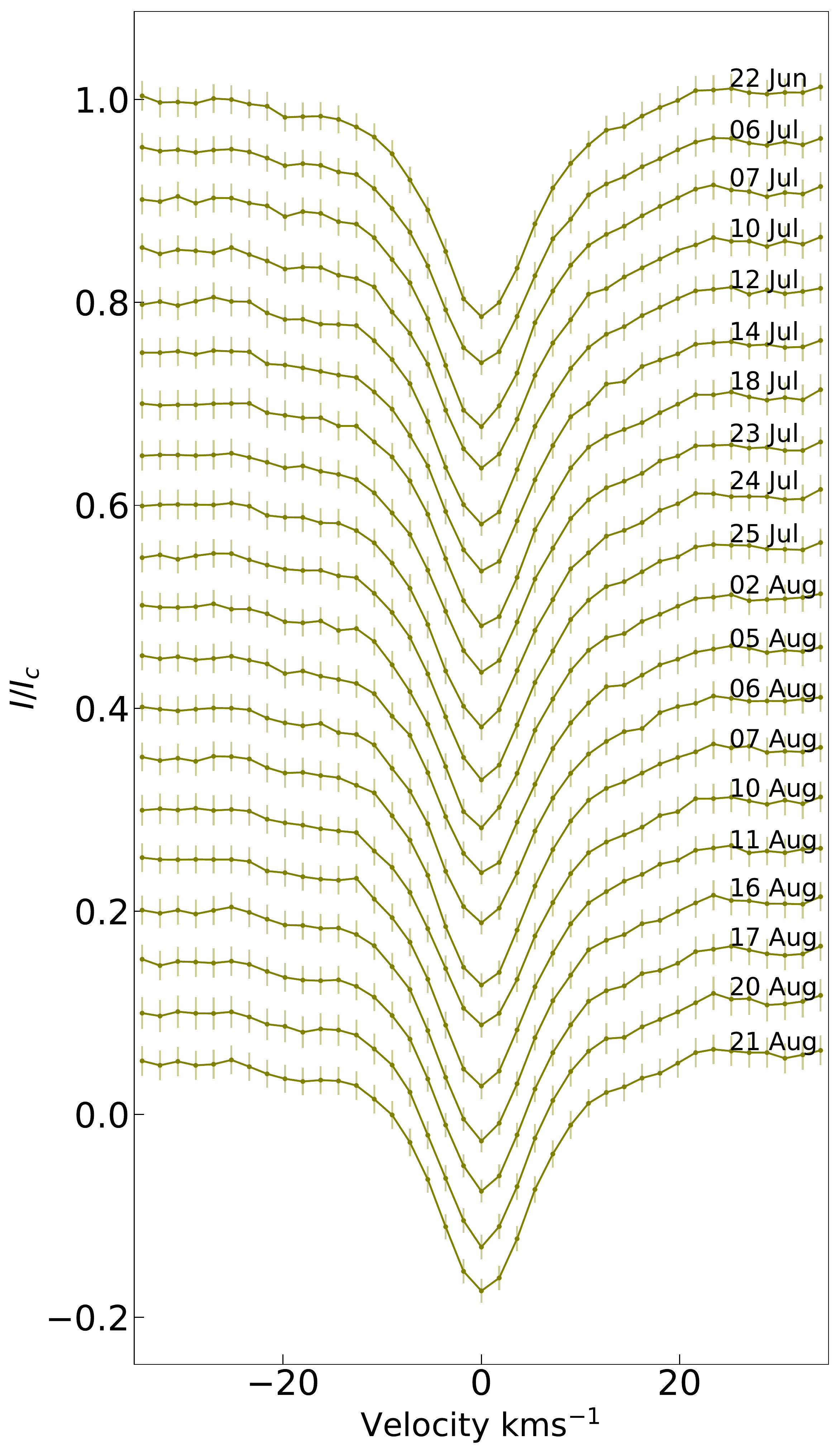}
    \includegraphics[width=0.3\linewidth]{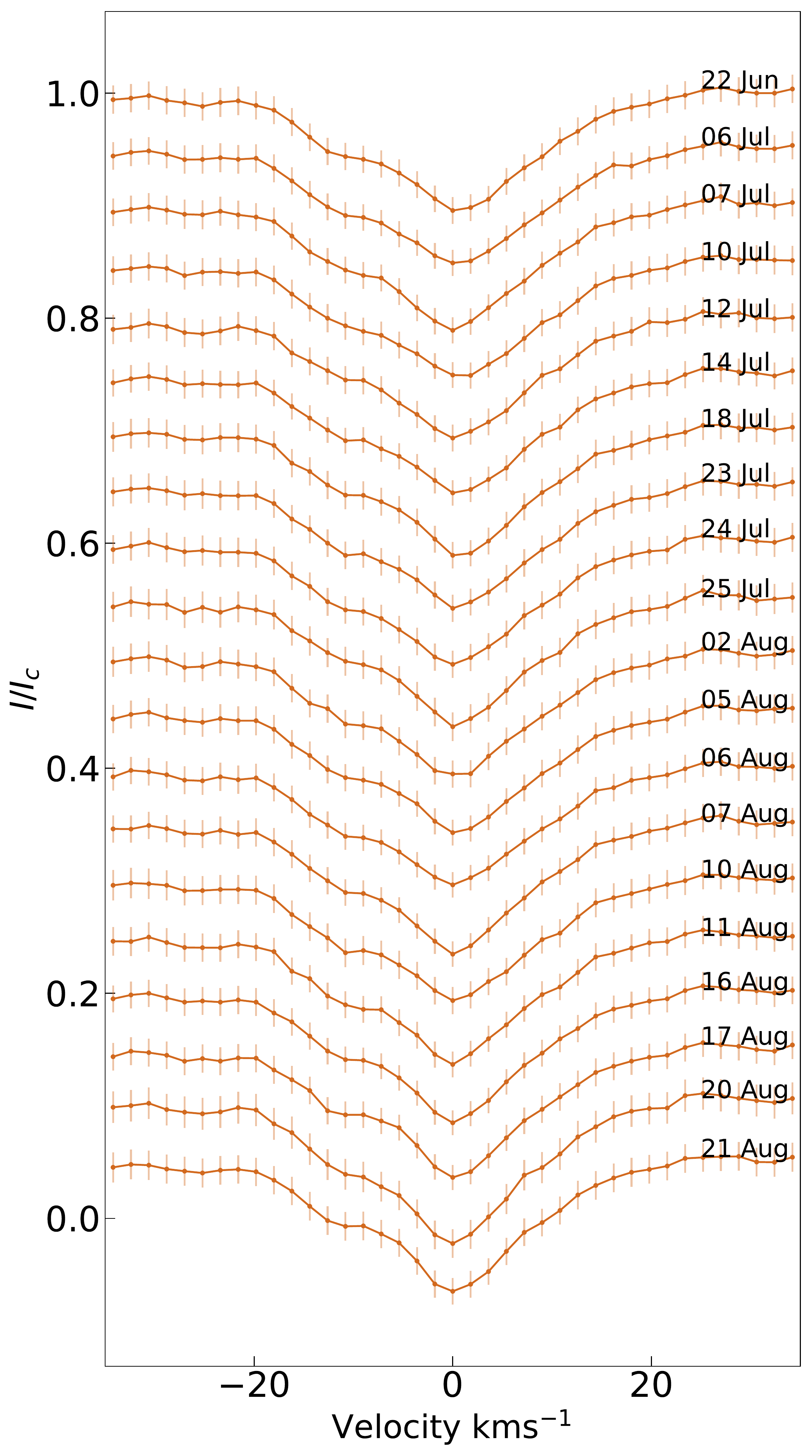}
    \caption{Time series of the Stokes $I$ profiles for the 2010 observations of EV Lac computed with (Left) the full mask without telluric window a priori, (Middle) union of the 2010-trained best sub-masks, and (Right) the 2010-trained worst RMS-based sub-mask without telluric windows a priori. The Stokes profiles are shifted vertically for visibility, and sorted with ascending observation date. We observe an increased sharpness when the union of best sub-masks is used, and a broad, distorted profile when the worst sub-mask is used, implying the presence of ``stable'' and ``unstable'' lines in the sub-masks, respectively. \label{fig:StokesI_submasks}}
\end{figure*}

We compare the distributions of the line parameters for three different masks: the full mask without telluric windows a priori (3240 lines), the 2010-trained union of best sub-masks (721 lines), and the 2010-trained worst RMS-based sub-mask without telluric windows a priori (2117 lines). The training refers to the 2010 data set of EV Lac with the randomised approach. 

Fig.\ref{fig:line_distribs} illustrates the distributions of line depth, wavelength and Land\'e factor for the three masks. The depth distribution for the full mask spans between 0.4 and 1.0 the continuum intensity, with a third of the lines deeper than 0.9 and with a mean of 0.73. The wavelength distribution features an almost exponential decay from 350 to 1080 nm and a mean of 433.1 nm, while the g$_\mathrm{eff}$ distribution ranges between 0.0 and 3.4, with a mean of 1.25 and two peaks at 1.0 and 1.5.

Apart from the different number of lines in the bins due to our randomised downselection approach, the two other sub-masks do not feature particular trends or quantitative differences in the statistics, implying that it is not possible to straightforwardly build ``stable'' or ``unstable'' sub-masks with a selection based on these parameters.

Despite the absence of a particular feature in the distributions, the effects on Stokes $I$ profiles of these masks are striking. From Fig.~\ref{fig:StokesI_submasks} we observe how the union of the best sub-masks leads to deeper and narrower profiles than the full mask, which would result in a more precise RV estimate. On the contrary, the application of the worst sub-mask yields shallow and distorted profiles, therefore increasing the dispersion of the associated RV data set. These features help us demonstrate further the capability of our algorithm to discern ``stable'' and ``unstable'' lines.

\end{appendix}

\end{document}